\documentclass[sigconf]{acmart}
\copyrightyear{2024}
\acmYear{2024}
\setcopyright{acmlicensed}\acmConference[CCS '24]{Proceedings of the 2024
ACM SIGSAC Conference on Computer and Communications Security}{October
14--18, 2024}{Salt Lake City, UT, USA}
\acmBooktitle{Proceedings of the 2024 ACM SIGSAC Conference on Computer
and Communications Security (CCS '24), October 14--18, 2024, Salt Lake City,
UT, USA}
\acmDOI{10.1145/3658644.3690316}
\acmISBN{979-8-4007-0636-3/24/10}

\AtBeginDocument{%
  }

\begin{document}

\title[Towards More Practical Membership Inference Attacks]{Is Difficulty Calibration All We Need? Towards More Practical Membership Inference Attacks}

\author{Yu He}
\authornote{Equal contribution.}
\authornote{The Key Laboratory of Aerospace Information Security and Trusted Computing, Ministry of Education, School of Cyber Science and Engineering, Wuhan University.}
\affiliation{
  \institution{Wuhan University}
  \city{Wuhan}
  \country{China}
}
\email{yuherin@whu.edu.cn}

\author{Boheng Li}
\authornotemark[1]
\authornotemark[2]
\affiliation{
  \institution{Wuhan University}
  \city{Wuhan}
  \country{China}
}
\email{randy.bh.li@foxmail.com}

\author{Yao Wang}
\authornotemark[2]
\affiliation{
  \institution{Wuhan University}
  \city{Wuhan}
  \country{China}
}
\email{valuewang@whu.edu.cn}

\author{Mengda Yang}
\authornotemark[2]
\affiliation{
  \institution{Wuhan University}
  \city{Wuhan}
  \country{China}
}
\email{mengday@whu.edu.cn}

\author{Juan Wang}
\authornote{Corresponding author.}
\authornotemark[2]
\affiliation{
  \institution{Wuhan University}
  \city{Wuhan}
  \country{China}
}
\email{jwang@whu.edu.cn}

\author{Hongxin Hu}
\affiliation{
  \institution{University at Buffalo}
  \city{Buffalo}
  \state{NY}
  \country{United States} 
}
\email{hongxinh@buffalo.edu}

\author{Xingyu Zhao}
\affiliation{
  \institution{University of Warwick}
  \city{Warwickshire}
  \country{UK}
}
\email{xingyu.zhao@warwick.ac.u}

\renewcommand{\shortauthors}{Yu He et al.}


\begin{abstract}
The vulnerability of machine learning models to Membership Inference Attacks (MIAs) has garnered considerable attention in recent years. These attacks determine whether a data sample belongs to the model's training set or not. Recent research has focused on reference-based attacks, which leverage difficulty calibration with independently trained reference models. While empirical studies have demonstrated its effectiveness, there is a notable gap in our understanding of the circumstances under which it succeeds or fails. In this paper, we take a further step towards a deeper understanding of the role of difficulty calibration. Our observations reveal inherent limitations in calibration methods, leading to the misclassification of non-members and suboptimal performance, particularly on high-loss samples. We further identify that these errors stem from an imperfect sampling of the potential distribution and a strong dependence of membership scores on the model parameters. By shedding light on these issues, we propose RAPID: a query-efficient and computation-efficient MIA that directly \textbf{R}e-lever\textbf{A}ges the original membershi\textbf{P} scores to m\textbf{I}tigate the errors in \textbf{D}ifficulty calibration. Our experimental results, spanning 9 datasets and 5 model architectures, demonstrate that RAPID outperforms previous state-of-the-art attacks (e.g., LiRA and Canary offline) across different metrics while remaining computationally efficient. Our observations and analysis challenge the current de facto paradigm of difficulty calibration in high-precision inference, encouraging greater attention to the persistent risks posed by MIAs in more practical scenarios.\footnote{Code is available at \href{https://github.com/T0hsakar1n/Is-Difficulty-Calibration-All-We-Need-Towards-More-Practical-Membership-Inference-Attacks}{https://github.com/T0hsakar1n/Is-Difficulty-Calibration-All-We-Need-Towards-More-Practical-Membership-Inference-Attacks}.}
\end{abstract}



\maketitle

\section{Introduction}
More personal privacy data has been incorporated into the datasets used for Machine Learning (ML) recently (such as medical~\cite{esteva2017dermatologist} and communication records~\cite{chen2019gmail}). It is thus important to investigate whether models can effectively prevent privacy leakage. Membership Inference Attacks (MIAs)~\cite{shokri2017membership} have been proposed to measure the extent of a model's leakage of member samples. It aims to predict whether a given data point belongs to the training set of a given target model or not. MIAs are now the de facto standard evaluation method for models' privacy risks~\cite{murakonda2020ml, song2020introducing} due to their simplicity to serve as a direct threat and the fact that MIAs are an important component of more sophisticated attacks~\cite{carlini2021extracting}.

Typically, MIAs exploit models' tendency to overfit their training data and therefore exhibit discrepancies in the outputs between members and non-members. Previous work seeks to learn these distinctive statistical features from the model’s original outputs in different ways, such as training a binary classifier (\textit{attack model})~\cite{shokri2017membership, leino2020stolen, nasr2019comprehensive, salem2018ml} or manually computing metrics like loss~\cite{yeom2018privacy} or entropy~\cite{song2021systematic}. While these attacks have demonstrated excellent performance on average-case metrics (Accuracy or ROC-AUC~\cite{sankararaman2009genomic}), Carlini et al.~\cite{carlini2022membership} point out that they do not pose important privacy risks: high Accuracy/AUC are mainly due to the successful identification of non-members rather than members. This limitation can be attributed to the fact that the influence of samples' intrinsic difficulty~\cite{carlini2022membership, sablayrolles2019white, watson2021importance} on the obtained model outputs is neglected. Specifically, certain simple non-member samples such as images with distinctive features or very short sentences, exhibit similarity to members in terms of model outputs, showing higher membership scores~\cite{Fan_Lewis_Dauphin_2018}.

Difficulty calibration is proposed by Watson et al.~\cite{watson2021importance} to mitigate the aforementioned issues. It attempts to quantify the difficulty of sample points (i.e., the extent to the sample represented on the whole distribution) and uses this value to regularize the model's original outputs, finally obtaining \textit{calibrated scores} for MIAs. In practice, difficulty calibration primarily measures the difficulty of target samples by feeding them into models trained on similar data (reference models). A category of attacks known as reference-based attacks employs this technique, aiming to achieve finer-grained calibration at the cost of extensive computational resources and numerous queries~\cite{sablayrolles2019white, carlini2022membership, ye2022enhanced, wen2022canary}.

While existing reference-based attacks have achieved significant breakthroughs on recently recommended True-Positive Rate at low False-Positive Rate (\textit{TPR at low FPR}), we argue that difficulty calibration is ``not all we need'' to achieve more powerful and practical MIAs. We have observed that some non-members, which could have been correctly classified, are inadvertently misclassified after difficulty calibration~\cite{watson2021importance}, leading to suboptimal performance. Typically, difficulty calibration assumes that outputs from reference models can effectively represent the difficulty of target samples. However, such an assumption is optimistic and thus unrealistic in many cases. In this paper, we examine and highlight that calibration errors primarily stem from two contributing factors: 1) the reference dataset is a subset sampled from the potential distribution; 2) membership scores are highly dependent on the model parameters. A more comprehensive analysis of this will be provided in Section~\ref{sec: 3}.

To effectively and efficiently address this issue, we propose RAPID that directly \textbf{R}e-lever\textbf{A}ges the original membershi\textbf{P} scores to m\textbf{I}tigate the errors in \textbf{D}ifficulty calibration, rather than treating it merely as a component of obtained calibrated scores. Specifically, while the original scores are strongly influenced by the inherent difficulty of the samples, they can provide reliable non-membership evidence because the target model directly fits member points during training. In other words, samples exhibiting extremely low original scores (e.g., high losses) are almost non-members. RAPID re-leverages the original outputs to correct misclassifications of non-members after difficulty calibration, thereby outperforming existing reference-based attacks. 

To mount RAPID, we adopt a typical supervised learning approach. Concretely, the adversary first trains a surrogate model (shadow model) for the target model and several reference models. Then, the adversary evaluates the shadow dataset samples' losses (or other signal outputs) on the shadow model and reference models to obtain original membership scores and calibrated scores. The adversary can use these two scores as features to train a \textit{scoring model}, which maps them to final scores for a threshold attack. In the end, the scoring model takes as input the target sample’s calibrated scores as well as its original scores to infer the sample’s membership status. The key point is that our approach introduces a \textbf{shortcut} in the inference from original outputs to membership status, allowing it to contribute independently. In contrast, previous work has been emphasizing the unreliability of original outputs and solely using them to serve as a component of calibrated scores~\cite{long2017towards, sablayrolles2019white, watson2021importance, carlini2022membership, ye2022enhanced, wen2022canary}. More importantly, RAPID eliminates the need for training a large number of reference models which is time-consuming, and it does not require near-unlimited query access to the target model, which is widely employed by existing state-of-the-art attacks~\cite{carlini2022membership, ye2022enhanced, liu2022membership, wen2022canary}. We leave a more detailed analysis of attack cost in Section ~\ref{sec: 3} and Section ~\ref{sec: 5}.

We conduct extensive experiments measuring the performance of our proposed RAPID, with comparisons to other
advanced attack methods. Experimental results show that RAPID achieves superior performance across various metrics while keeping its practicality in real-world scenarios. Notably, RAPID is able to achieve 5.1\% TPR at 0.1\% FPR on the CIFAR-10 dataset, approximately 2.5 times and 3 times the performance of state-of-the-art attacks,  LiRA offline~\cite{carlini2022membership} and Canary offline~\cite{wen2022canary}. Furthermore, RAPID shows a relative improvement of 21.4\% in AUC and 14.6\% in Balanced Accuracy. All improvements are achieved with only 1/25 of LiRA offline's computational cost (and potentially lower). To make stronger conclusive statements regarding the practicality of RAPID, we also evaluate it in the realm of Large Language Models (LLMs), which has seen limited exploration in prior research. Experimental results show that RAPID can achieve approximately 3 times the TPR at 0.1\% FPR of attacks that only employ difficulty calibration on BERT~\cite{devlin2018BERT}. We conduct extensive ablation studies to evaluate the influence of various components on our attack, such as the number of queries, the number of reference models, as well as the level of knowledge regarding model architecture and data distribution of the adversary. Finally, we provide additional discussion on the advantages of directly re-leveraging the original membership scores by introducing our shortcut in more complicated reference-based attacks, and fairly comparing RAPID with the most powerful (though computationally infeasible) LiRA online version. We also highlight the limitations of our work, which could provide potential directions for future research. In summary, our paper makes the following contributions:

\begin{itemize}\setlength{\itemsep}{0.2em}
\item We discover and analyze the phenomena that inherent errors in difficulty calibration may lead to the misclassification of non-members, who could have otherwise been accurately classified, resulting in suboptimal performance.
\item We propose a straightforward yet powerful MIA, known as RAPID, to address the errors in difficulty calibration, successfully outperforming other state-of-the-art attacks and being arguably more practical.
\item We conduct extensive experiments in both the classic image domains and the recent field of LLMs to demonstrate the effectiveness and efficiency of our attack.
\end{itemize}
\section{Background}
\label{sec: 2}

\subsection{Machine Learning}
A learned neural network in machine learning classification tasks can be represented as a parameterized function \(\mathcal{M}_{\text{\(\theta\)}}: \mathcal{X}\rightarrow\mathcal{Y}\) that maps each input \(\mathit{x}\in\mathcal{X}\) to a probability vector over a group of class labels \(\mathcal{Y}\). \(\theta\) is the set of parameters. To obtain the optimal weights \(\theta\), we utilize a dataset \(\mathcal{D}\) sampled from an underlying distribution \(\pi\). The process of learning the neural network model is denoted as \(\mathcal{M}_\theta\leftarrow\mathcal{T}(\mathcal{D})\), where the training algorithm \(\mathcal{T}\) is applied to the training set \(\mathcal{D}\) to optimize the weights \(\theta\). The training process is performed by minimizing the empirical loss using the stochastic gradient descent algorithm:
\begin{equation}
\theta_{i+1}\leftarrow\theta_i-\epsilon\sum_{(x,y)\in\mathcal B}\nabla_\theta\mathcal L(\mathit y,\mathcal M_{\theta_i}(\mathit x)),
\label{eq:(1)}
\end{equation}
where \(\mathcal{B}\) is a small batch of training samples, \(\epsilon\) the learning rate for iteratively updating the parameters \(\theta\) of the neural network and \(\mathcal{L}\) the prediction loss such as cross-entropy loss. Utilizing the gradient descent from Equation \ref{eq:(1)} will inevitably drive the training sample's loss \(\mathcal{L}\)(\(\mathit{y}\),\(\mathit{p}\)), \(\mathit{x}\in\mathcal{D}\) to zero. However, achieving strong generalization to unseen dataset \(\mathcal{D}_\text{test}\in\pi\) remains a challenge in neural network training. Data augmentation~\cite{zhong2020random, van2001art, cubuk2018autoaugment} serves as an effective technique to significantly improve test accuracy. In order to make the attack scenario more realistic, we also apply this technique to the training of the target model.

\subsection{Membership Inference Attacks}
MIAs aim to infer whether a sample belongs to the training set of a victim model. It has drawn much attention~\cite{carlini2021extracting, chen2020gan, hayes2017logan, mcmahan2018general, nasr2019comprehensive, song2019auditing, zhang2021membership, he2021node, he2022membership,wuyou} because of its direct threats to privacy and its ease of deployment. Membership inference is also widely used to measure the effectiveness of machine unlearning \cite{bourtoule2021machine} and serve as a baseline for data tracing \cite{wang2024did,wangtrace} and ownership verification \cite{shao2025explanation,Huang_2023_ICCV}.

\paragraph{Definition of MIAs.} Let \(\pi\) be the underlying distribution, let \(\mathcal{T}\) be the training algorithm that a challenger (defender) would use, and let \(\mathcal{A}\) be the attack method an attacker would use to make a prediction. The game will proceed as follows:
\begin{enumerate}
    \item The challenger samples a training dataset \(\mathcal{D}\in\pi\) and trains a target model \(\mathcal{M}_\theta\leftarrow\mathcal{T}(\mathcal{D})\).
    \item The challenger randomly flips an unbiased coin \(\mathit{b}\in\phantom{1}\{0,1\}\).
    \item If \(\mathit{b} = 0\), the challenger randomly samples a fresh target point \(\mathit{x}\in\pi\setminus\mathcal{D}\). Otherwise, the challenger randomly samples a fresh target point \(\mathit{x}\in\mathcal{D}\).
    \item The challenger sends \((\mathit{x},\mathit{y})\) to the attacker. The attacker has black-box access to the model \(\mathcal{M}_\theta\) and the distribution \(\pi\setminus\mathcal{D}\).
    \item The challenger outputs a bit \(\hat{\mathit{b}}\leftarrow\mathcal{A}_{\mathcal{M}_\theta}^{\pi\setminus\mathcal{D}}(x,y)\).
    \item if \(\hat{\mathit{b}} = \mathit{b}\), output 1. Otherwise, output 0.
\end{enumerate}
It is worth noting that Yeom et al.~\cite{yeom2018privacy} also made a similar definition to ours. However, they assume the attacker has access to the entire distribution \(\pi\), which means they can access potential training data before inference time. This assumption probably favors the attacker and is subject to being unrealistic. Therefore, we have modified the definition to restrict the attacker's access to only the attack dataset \(\mathcal{D}_\text{attack}=\pi\setminus\mathcal{D}\). This modification is for assessing the attack method on unseen data. As the attack dataset \(\mathcal{D}_\text{attack}\) and the target model \(\mathcal{M}_\theta\) are typically considered fixed, the adversary’s prediction can be simplified as \(\mathcal{A}(x,y)\).

\paragraph{Attack Method.} To provide a better understanding of MIAs, we make a formal definition of \(\mathcal{A}\) as follows:
\begin{equation}
\mathcal{A}(x,y)=\mathbbm{1}[\mathcal{S}(x,y)>t],
\label{eq:(2)}
\end{equation}
where \(\mathcal{S}\) outputs the membership score of \((x,y)\) and \(\mathit{t}\) indicates a threshold used for decision-making. For illustrative purposes, we start out by using the loss value~\cite{yeom2018privacy} to represent the membership score \(\mathcal{S}(x,y)\), so that \(\mathcal{A}_\text{loss}(x,y)=\mathbbm{1}[-\mathcal{L}(y,\mathcal{M}_\theta(x))>t]\). This is intuitive as machine learning models are trained to minimize the loss of their members. Consequently, members naturally have smaller losses compared to non-members, which can be used to distinguish between them. Prior work~\cite{sablayrolles2019white, ye2022enhanced} has also proved that loss-threshold-based attacks are theoretically powerful. The subsequent work has built upon this and focused on high-precision membership inference~\cite{watson2021importance,sablayrolles2019white,ye2022enhanced,liu2022membership,carlini2022membership,wen2022canary} or altering the assumptions about the attacker's knowledge (i.e., attack scenario)~\cite{nasr2019comprehensive,leino2020stolen,choquette2021label,li2021membership}.

\paragraph{Metrics.} We consider the following three common metrics:
\begin{itemize}\setlength{\itemsep}{0.1em} \setlength{\parskip}{0.1em}
\item \textbf{Balanced Accuracy.} The simplest method to evaluate attack efficacy that measures how often an attack correctly predicts membership on a balanced dataset of members and non-members~\cite{choquette2021label, hayes2017logan, leino2020stolen, nasr2019comprehensive, sablayrolles2019white, song2021systematic, truex2018towards, yeom2018privacy, liu2022membership}.
\item \textbf{AUC.} The most commonly used method to interpret the Receiver Operating Characteristic (ROC) curve~\cite{sankararaman2009genomic} is by calculating the area under the curve (AUC). It reflects the average-case success of membership inference.
\item \textbf{TPR at Low FPR.} The latest metric used to evaluate attack efficacy focuses on the TPR of the attack when the threshold \(t\) is set to a large value to achieve an extremely low FPR. This metric is currently recommended~\cite{carlini2022membership} because it directly reflects the actual extent of privacy leakage of the model towards its training samples. 
\end{itemize}

\begin{figure*}[h]
  \centering
  \includegraphics[width=\linewidth]{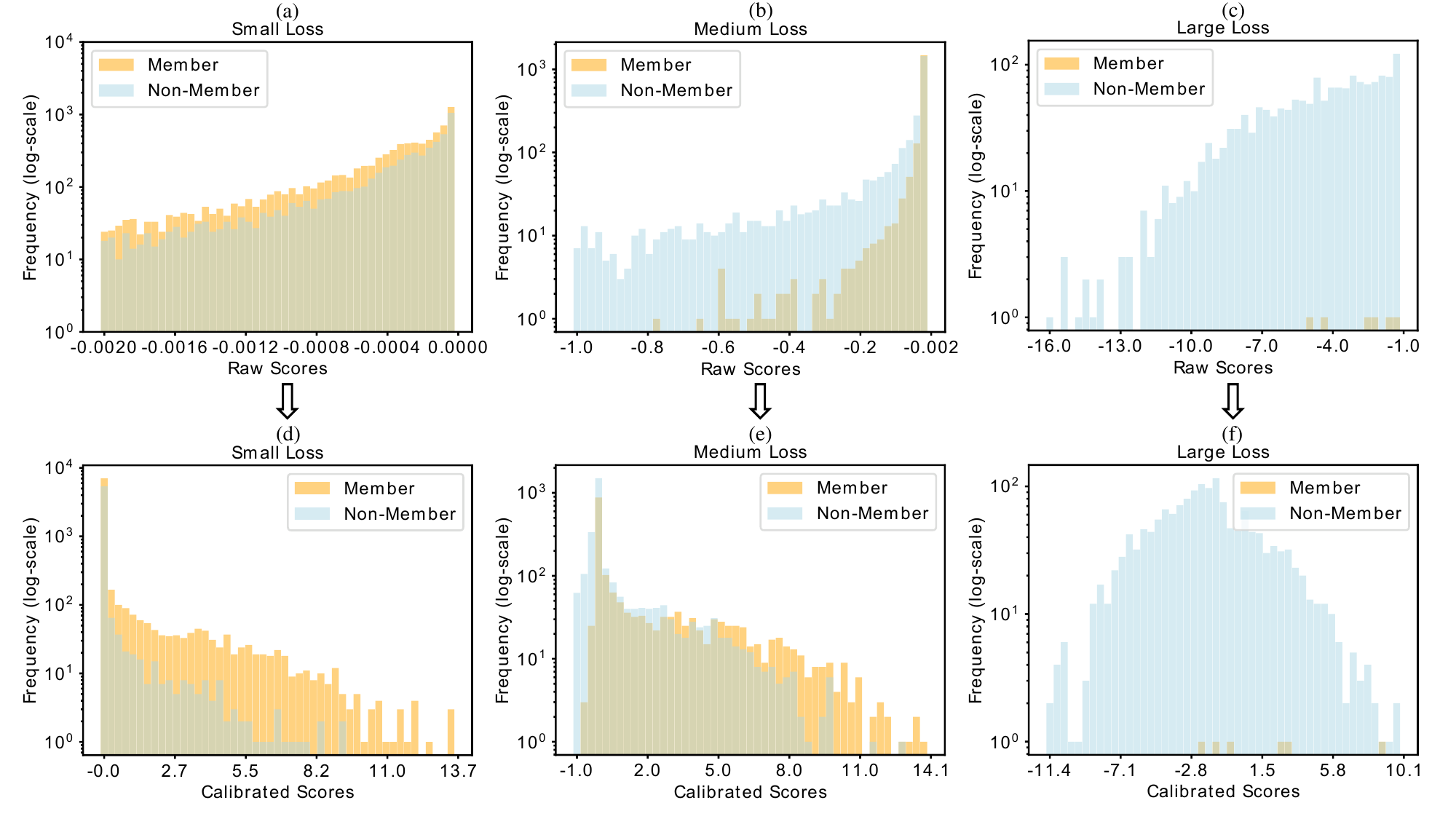}
  \caption{The distribution of raw membership scores and calibrated membership scores. All the samples with different losses obtained from the target VGG16 model are divided into three ranges: `small loss'[0,0.002), `medium loss'[0.002,1), and `large loss'[1,$\infty$). The target model is trained on the CIFAR-10 dataset. Difficulty calibration significantly increases the membership scores of some non-member samples that originally had medium or large losses.}
  \label{Figure: vanilla-calibration}
\end{figure*}

\section{Rethinking Difficulty Calibration}
\label{sec: 3}
In this section, we provide a detailed rethinking of difficulty calibration~\cite{watson2021importance}, which forms the basis for a variety of advanced attacks. We first provide additional background on difficulty calibration. Then, we conduct an in-depth analysis of its limitations. Finally, we present the design intuition of RAPID.
 
\paragraph{Difficulty Calibration.} Watson et al.~\cite{watson2021importance} have argued that \(\mathcal{A}_\text{loss}\) is very unreliable as samples have different intrinsic difficulty. A sample exhibiting low loss is not necessarily a member; it could also be due to its low difficulty. \(\mathcal{A}_\text{loss}\) struggle to separate these low-loss non-members from typical members since both can attain a high membership score. To improve the attack's reliability, a simple modification to the original membership score called difficulty calibration is required. It is based on the intuition that if \((x,y)\) has low difficulty (i.e., over-represented on \(\pi\)), it will generally show high membership scores on all reference models trained on data similar to that of the target model. By subtracting the average of membership scores on reference models from the original membership score of the target sample, the influence of sample difficulty can be eliminated. Formally, the calibrated membership scores can be defined as:
\begin{equation}
\mathcal{S}'(x,y)=\mathcal{S}(x,y)-\mathbbm{E}_{\mathcal{M}_{\text{ref}}\leftarrow\mathcal{T}(\mathcal{D}_{\text{attack}})}[\mathcal{S}(x,y)],
\label{eq:(3)}
\end{equation}
where the expectation is approximated by sampling several reference models from \(\mathbbm{T}(\mathcal{D}_{\text{attack}})\). From Equation \ref{eq:(3)} we can find that ideally, calibrated scores of non-members will approximate 0, as their original scores are mainly up to their intrinsic difficulty. Conversely, the calibrated scores of members will exceed 0 because their original scores are influenced not only by their difficulty but also by the training process itself. This technique, which we called difficulty calibration, has demonstrated breakthroughs in high-precision attacks and served as the foundation for subsequent advanced reference-based attacks~\cite{sablayrolles2019white, carlini2022membership, ye2022enhanced, wen2022canary}.

\paragraph{Limitations.} To understand the circumstances under which the difficulty calibration succeeds or fails, we divide the samples with different losses obtained from the target model into three ranges: `small loss'[0,0.002), `medium loss'[0.002,1), and `large loss'[1,$\infty$). Specifically, Figure \ref{Figure: vanilla-calibration}(a), Figure \ref{Figure: vanilla-calibration}(b), and Figure \ref{Figure: vanilla-calibration}(c) categorize samples within specific ranges of losses, while Figure \ref{Figure: vanilla-calibration}(d), Figure \ref{Figure: vanilla-calibration}(e), and Figure \ref{Figure: vanilla-calibration}(f) represent the frequency distributions of calibrated scores for these samples corresponding to Figure \ref{Figure: vanilla-calibration}(a), Figure \ref{Figure: vanilla-calibration}(b), and Figure \ref{Figure: vanilla-calibration}(c), respectively. To dispel potential misunderstandings, we emphasize that calibrated score distributions (i.e., the X-axis) of (d), (e), and (f) may intersect, and the different scales of the Y-axis are due to the different number of points included in them. Comparing (a) to (d), the calibrated signal indeed allows for scored highest samples to belong to the member class, making it possible to confidently predict member samples at low FPR. Therefore, if we only consider distinguishing between members/over-represented non-members, difficulty calibration indeed performs exceptionally well. However, we can observe that medium-loss and large-loss non-members, which could have been classified correctly, have a larger overlap with members in the distribution of scores after difficulty calibration. Specifically, the increased overlap area from (b) to (e) may cause degradation in metrics reflecting the average privacy leakage~\cite{watson2021importance}, such as Balanced Accuracy and AUC. Furthermore, the shift from (c) to (f) highlights an issue where half of the non-member samples witness a surge in their membership scores, with some even surpassing 7. High calibrated scores of large-loss non-members may render the selection of an appropriate threshold \(t\) more challenging, as the crucial metric TPR at low FPR is highly sensitive to non-members with high membership scores~\cite{carlini2022membership}. Overall, Figure~\ref{Figure: vanilla-calibration} shows that depending solely on difficulty calibration constitutes a suboptimal approach, with respect to both average-case metrics and TPR at low FPR.

These limitations arise from two main factors. Firstly, the average results of membership scores obtained from reference models cannot precisely depict the extent to the target record \((x,y)\) represented within the distribution \(\pi\), as the \(\mathcal{D}_\text{attack}\) only represents a subset of distribution \(\pi\). There is an inherent difference between these two distributions. For instance, a target sample may be over-represented on a subset sampled from \(\pi\) but not in the entire distribution. This error can be partially mitigated by conducting multiple random samplings from \(\mathcal{D}_\text{attack}\), provided that the attacker possesses a \(\mathcal{D}_\text{attack}\) larger than the target model's training set. A larger \(\mathcal{D}_\text{attack}\) implies a better approximation to the true distribution. Secondly, the calibrated scores of each sample heavily depend on the parameters of the target model and the reference models. To better illustrate this, consider the following distribution: \(\mathbbm{S}(x,y)=\{-\mathcal{L}(y,\mathcal{M}(x)\leftarrow\mathcal{T}(\mathcal{D}))\mid \mathcal{T}\in\mathbbm{T}\}\) is the
distribution of losses on \((x,y)\) for models trained on a given dataset using different training algorithms. We follow previous work~\cite{carlini2022membership} to model \(\mathbbm{S}\) as a Gaussian distribution:
\begin{equation}
\mathbbm{S}(x,y) \sim \mathcal{N}(\mu, \sigma^2).
\end{equation}
For simplicity, we make an assumption that the distribution of losses on \((x,y)\) for the target model and the reference model are independent of each other. By calculating the difference of two independent Gaussian distributions, i.e., \(\mathbbm{S}_\text{tar}(x,y)\) and \(\mathbbm{S}_\text{ref}(x,y)\), we can quantify the distribution of target samples' calibrated membership scores as:
\begin{equation}
\mathbbm{S}_\text{cal}(x,y) \sim \mathcal{N}(\mu_\text{tar}-\mu_\text{ref}, \sigma^2_\text{tar}+\sigma^2_\text{ref}),
\end{equation}
where \(\mu_\text{tar}\), \(\mu_\text{ref}\), \(\sigma^2_\text{tar}\) and \(\sigma^2_\text{ref}\) are uniquely determined by the target record \((x,y)\) and given training sets. In a single security game, the specific parameters of the target model and the reference model actually represent a single random sampling from the distribution \(\mathbbm{S}_\text{cal}(x,y)\). Therefore, the calibrated scores depend significantly on the parameters of models and not just on membership status. For non-members, \(\mu_\text{tar}\) and \(\mu_\text{ref}\) should behave similarly statistically. This would make the mean of the distribution \(\mathbbm{S}_\text{cal}\) close to 0, ideally. However, the increased variance leads to a significant occurrence of calibrated scores much larger than 0, resulting in the misclassification of non-members. In worse cases, since \(\mathcal{D}_\text{attack}\) and the target model training set do not intersect at all, it may lead to \(\mu_\text{tar}\) being noticeably larger than \(\mu_\text{ref}\) for non-members.

\paragraph{Design Intuition.} Machine learning algorithms aim to minimize the loss during training, which implies that high original loss can provide sufficient evidence of non-members. On the other hand, difficulty calibration indeed helps confidently separate low-loss non-members from members. However, the aforementioned errors may cause an unexpected increase in membership scores of some high-loss non-members, making this attack suboptimal. We can thus utilize the original membership scores to differentiate between non-members who scored higher due to the difficulty of calibration and genuine members. Our method does not require training numerous reference models~\cite{carlini2022membership, ye2022enhanced, wen2022canary} to conduct a parametric likelihood-ratio test or querying the target model extensively~\cite{ye2022enhanced, liu2022membership} to mitigate the influence of target model parameters. By introducing a \textbf{shortcut} in the inference from original outputs to membership status, our proposed RAPID can significantly (as shown in our experiments) enhance the performance of MIAs while ensuring their practicality.

\begin{figure}[t]
    \centering
    \includegraphics[width=0.47\textwidth]{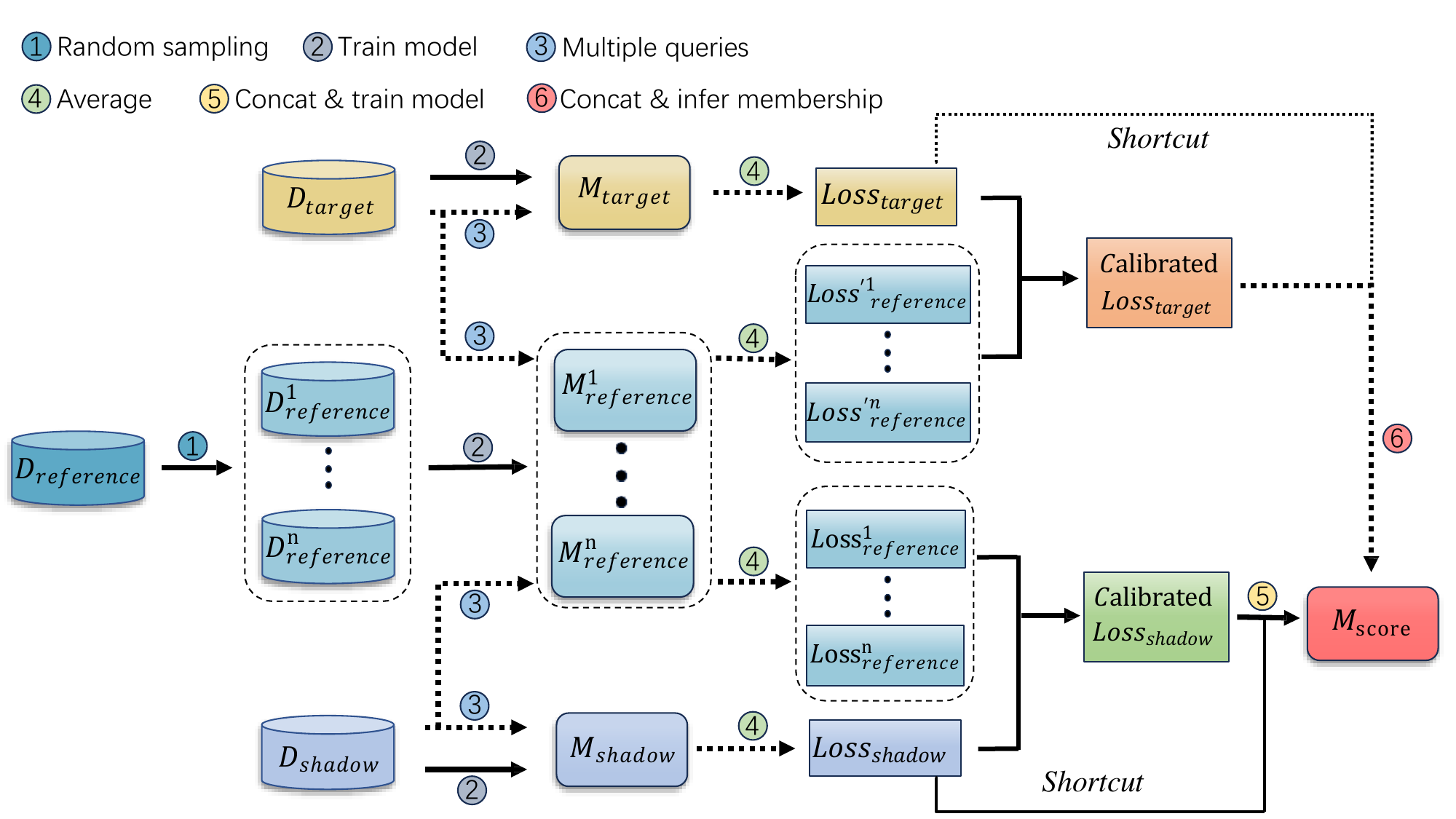}
    \caption{General attack pipeline of our RAPID.}
    \label{Figure pipline}
\end{figure}

\section{Attack Methodology}
\label{sec: 4}
In this section, we present the methodology of RAPID. We begin by defining the threat model that our attack operates under. Then we outline the pipeline of RAPID when the attacker has only black-box access to the target model. Finally, we introduce some useful techniques for enhancing the attack performance.

\subsection{Threat Model} In this paper, we primarily focus on the most commonly adopted black-box setting, in which the attacker only has access to the posterior probability distribution of the target model's outputs. We also follow previous advanced works~\cite{shokri2017membership, leino2020stolen, nasr2019comprehensive, salem2018ml, carlini2022membership, sablayrolles2019white, watson2021importance, liu2022membership, ye2022enhanced, wen2022canary}, assuming that the attacker can sample sufficient data from \(\pi\setminus\mathcal{D}\) and knows the target model's architecture. We will show these two assumptions can be relaxed in Section ~\ref{sec: 5}. Recently, there have been extensions of MIAs from black-box scenarios to settings such as white-box scenarios~\cite{nasr2019comprehensive, leino2020stolen} and label-only~\cite{choquette2021label,li2021membership}, which will not be discussed in this paper.

\subsection{Attack Method} We suggest training a scoring model \(\mathcal{M}_\text{score}\) to map the original membership scores \(\mathcal{S}(x,y)\) and the calibrated membership scores \(\mathcal{S}'(x,y)\) together to the final membership scores, which are then used for membership inference. Therefore, the definition in Equation (\ref{eq:(2)}) can be modified as follows:
\begin{equation}
\mathcal{A}(x,y)=\mathbbm{1}[\mathcal{M}_\text{score}(\mathcal{S},\mathcal{S'})>t].
\end{equation}
We train a model \(\mathcal{M}_\text{score}\) to find the optimal mapping toward final membership scores because original membership scores and calibrated membership scores have different scales. The scoring model learns to directly correct the prediction errors caused by the aforementioned errors in \(\mathcal{S}'(x,y)\) using \(\mathcal{S}(x,y)\). Specifically, instances with high calibrated membership scores but low original membership scores are expected to be non-members rather than members. Using a heuristic search algorithm to obtain the parameters required for optimal mapping is apparently suitable in this case. To conduct our attack, the adversary needs to perform four steps: shadow model training, reference model training, scoring model training, and membership inference. We give the detailed pipeline of our proposed RAPID in Figure \ref{Figure pipline}.

\paragraph{Shadow Model Training.} As the attacker does not have access to the target model's training dataset \(\mathcal{D}_\text{target}\). We thus sample a subset \(\mathcal{D}_\text{shadow}\)
 from \(\mathcal{D}_\text{target}\)'s i.i.d. (independent identically distributed) dataset \(\mathcal{D}_\text{attack}\) to train the shadow model \(\mathcal{M}_\text{shadow}\). It shares similarities with the target model \(\mathcal{M}_\text{target}\) in terms of properties, and we can utilize its outputs and \(\mathcal{D}_\text{shadow}\) to train \(\mathcal{M}_\text{score}\).

\paragraph{Reference Model Training.} The attacker then uses another subset of \(\mathcal{D}_\text{attack}\), referred to as \(\mathcal{D}_\text{reference}\) to train reference models \(\mathcal{M}_\text{reference}\). Different training algorithms \(\mathcal{T}\) can be used to obtain reference models with different parameters. See more detailed discussion in Section \ref{sec: 4.3}.

\begin{figure}[t]
    \centering
    \begin{subfigure}{0.23\textwidth}
        \includegraphics[width=\linewidth]{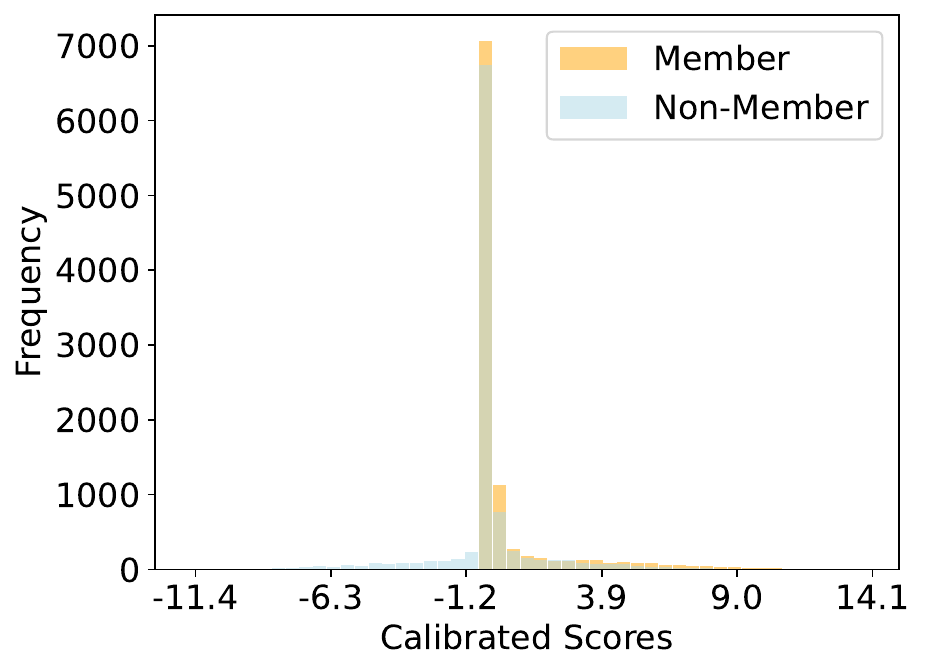}
    \end{subfigure}
    \hfill
    \begin{subfigure}{0.23\textwidth}
        \includegraphics[width=\linewidth]{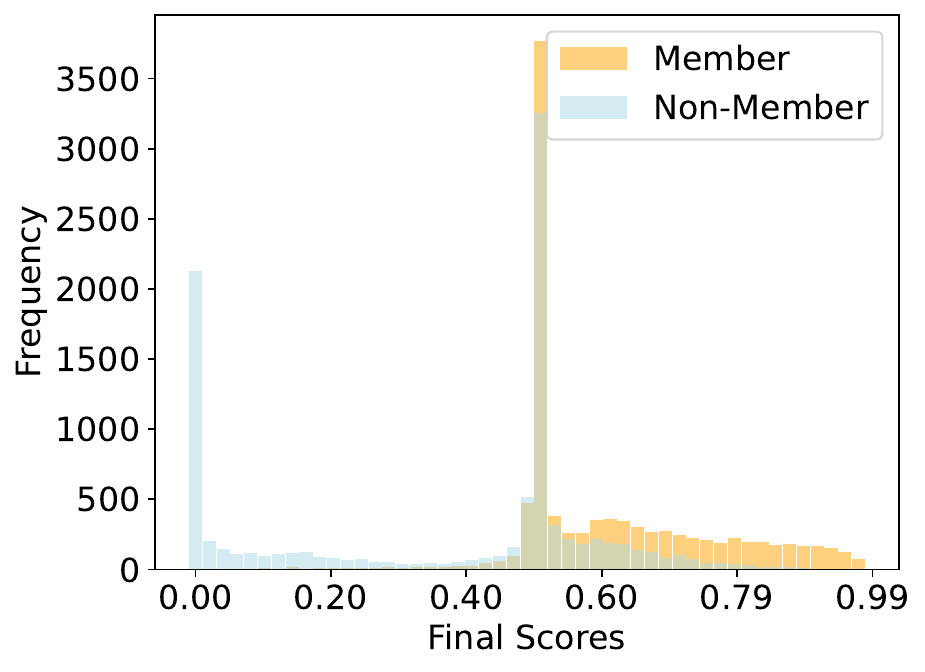}
    \end{subfigure}
    \caption{The frequency distributions of final scores and calibrated scores, which were sampled using a VGG16 model trained on CIFAR-10.}
    \label{Figure: final&calibrated}
\end{figure}

\paragraph{Scoring Model Training.} The attacker trains a scoring model, namely \(\mathcal{M}_\text{score}\), using the attack dataset \(\mathcal{D}_\text{attack}\). The scoring model is modeled as a Multi-Layer Perceptron (MLP) with a single output channel. To confine the output within the range of [0,1], a sigmoid layer is applied to the model's output. \(\mathcal{M}_\text{score}\) is thus defined as follows:
\begin{equation}
\mathcal{M}_\text{score}=sigmoid(\mathcal{MLP}(\mathcal{S}\oplus\mathcal{S'})).
\end{equation}
The model takes as input the concatenation of the original membership scores and the calibrated membership scores of the samples, while the corresponding labels are set to 1 if the sample belongs to the training set of the shadow model, and 0 otherwise. Binary Cross-Entropy Loss is utilized to compute the loss, and the objective is to minimize \(\mathcal{L}(\mathcal{M}_\text{score}(\mathcal{S},\mathcal{S'}),label)\) during training. Figure \ref{Figure: final&calibrated} demonstrates that the final membership scores obtained from \(\mathcal{M}_\text{score}\) exhibit a higher level of discrimination between members and non-members compared to the calibrated membership scores. It is worth noting that Yuan et al.~\cite{yuan2022membership} proposed a self-attention-based attack that utilizes the transformer~\cite{vaswani2017attention} to capture global dependencies among inputs and enables interaction within the inputs. We also experiment with modeling the scoring model as a transformer, but the results show no significant improvement. We believe this is because mapping the original scores and the calibrated scores to the optimal final scores is a simple task that an MLP can perform well. Future work will be conducted to further investigate the specific impact of the scoring model's architecture on attack performance.

\paragraph{Performing Membership Inference.} The attacker is finally able to conduct MIAs on each given sample. By feeding the target sample to both \(\mathcal{M}_\text{target}\) and \(\mathcal{M}_\text{reference}\), the attacker obtains \(\mathcal{S}(x,y)\) and \(\mathcal{S'}(x,y)\) respectively. These scores are concatenated and then input into \(\mathcal{M}_\text{score}\) to obtain the final membership score. To achieve optimal attack accuracy, the attacker simply needs to set the threshold \(t\) to 0.5. Through sweeping over a range of values for the threshold \(t\), the adversary can obtain the tradeoff between FPR and TPR, allowing for the calculation of AUC and the attack's TPR at a given low FPR. Compared to prior works, our attack method has an additional advantage: previous reference-based attacks do not provide guidance on the appropriate threshold \(t\) to achieve the desired low FPR attack in real scenarios~\cite{watson2021importance, sablayrolles2019white, carlini2022membership, long2020pragmatic, wen2022canary}. However, in our method, the shadow model can be utilized to determine an appropriate \(t\) in order to achieve the target FPR. 

\subsection{Generic Techniques}
\label{sec: 4.3}
 We will introduce two techniques used in our complete attack that significantly enhance the attack performance. We argue that these techniques could potentially serve as generic building blocks to enhance a variety of reference-based attacks.
\paragraph{Random Sampling.} As the size of \(\mathcal{D}_\text{attack}\) increases, the extent to the target record represented in the distribution \(\pi\) can be approximated better using \(\mathbbm{E}_{\mathcal{M}_{\text{ref}}\leftarrow\mathcal{T}(\mathcal{D}_{\text{attack}})}[\mathcal{S}(x,y)]\). Specifically, when the adversary possesses \(\mathcal{D}_\text{attack}\) significantly larger than \(\mathcal{D}_\text{target}\), the attack can be improved by training several \(\mathcal{M}_\text{reference}\) on different subsets of \(\mathcal{D}_\text{attack}\) sampled each time randomly. In more realistic scenarios where the attacker's available data is insufficient for random sampling, a substitute approach to improve the attack performance is by training multiple reference models using different training algorithms \(\mathcal{T}\), such as varying initialization parameters. This works as it brings the computed inherent difficulty (i.e., the average outputs) of the target record closer to \(\mu_\text{ref}\), thereby partially mitigating the impact resulting from the randomness in sampling from a distribution \(\mathbbm{S}_\text{ref}\).

\paragraph{Multiple Queries.} To make the observed original membership score of the target record closer to its \(\mu_\text{tar}\), a naive idea is to train multiple \(\mathcal{M}_\text{target}\) on \(\mathcal{D}_\text{target}\) and performs a single query on each of them. However, this appears to be an unfeasible strategy for the adversary. We can thus alternatively enhance the attack by averaging the outputs obtained from multiple queries on the same model on the target record and its augmentations. Notably, Carlini et al.~\cite{carlini2022membership} also apply this method to fit multiple-dimensional spherical Gaussians. They argue that these perturbed inputs may be seen by the target model during training and thus contain additional membership signals. However, we observe that simply averaging the outputs can already greatly enhance the attack, which is not entirely consistent with the explanation provided by Carlini et al.. We believe that this enhancement comes from mitigating the errors caused by the dependence of the target point's membership score on the parameters \(\theta\) of the target model. We leave a detailed discussion in Section~\ref{sec: 6.2}. As the method of fitting multiple-dimensional spherical Gaussians applies only to attacks using Gaussian likelihood estimate~\cite{carlini2022membership,wen2022canary}, we argue that subsequent MIAs can use the average membership scores obtained from multiple queries to replace the original scores to enhance their performance.

\begin{table*}[t]
\centering
\setlength{\tabcolsep}{3pt}
\caption{The prediction accuracy of different model architectures on different datasets.}
\scalebox{1.0}{
\begin{tabular}{l|cccccccc}
\toprule
Dataset&  \multicolumn{2}{c}{CIFAR-10}&  \multicolumn{2}{c}{CIFAR-100}& \multicolumn{2}{c}{CINIC-10}& \multicolumn{2}{c}{SVHN}\\
Model&Train acc&Test acc&Train acc&Test acc&Train acc&Test acc&Train acc&Test acc\\
\midrule
MobileNetV2&  {99.8\%}& 84.1\%&  {100.0\%}& 55.1\%&  {94.5\%}& 79.7\%&  {100.0\%}& 95.2\%\\
VGG16&  {99.8\%}& 82.4\%&  {99.9\%}& 48.5\%&  {99.9\%}& 80.0\%&  {100.0\%}& 94.7\%\\
ResNet50&  {98.1\%}& 75.0\%&  {100.0\%}& 41.1\%&  {99.8\%}& 79.8\%&  {99.8\%}& 94.2\%\\
DenseNet121&  {100.0\%}& 81.6\%&  {100.0\%}& 44.9\%&  {100.0\%}& 80.2\%&  {100.0\%}& 94.9\%\\
\bottomrule
\end{tabular}
}
\label{table:target model performances}
\end{table*}

\section{Evaluation}
\label{sec: 5}
In this section, we evaluate our RAPID on various benchmark datasets and diverse model architectures. We focus on three standard metrics: Balanced Acc, AUC, and TPR at low FPR, which have been detailed in Section \ref{sec: 2}. Through extensive experiments, we demonstrate that our attack outperforms the state-of-the-art methods and has lower attack costs. We also evaluate our attack under the defense of Differential Privacy (DP), which is a widely applied defense mechanism against privacy leakage attacks. In addition to the standard evaluation work emphasized by existing MIAs, we also conduct evaluations in the field of LLMs to explore the practicality of our attack. To distinguish these results from the classic evaluations, the specific experimental setup and results for this section are presented in Section~\ref{sec: 5.3}.

\subsection{Experimental Setup}
\label{sec: 5.1}
\paragraph{Datasets.} In the main experimental section, we select four benchmark image datasets, namely CIFAR-10~\cite{krizhevsky2009cifar} (a benchmark dataset used for classification tasks), CINIC-10~\cite{darlow2018cinic} (an extension of CIFAR-10 consisting of 270,000 images, with downsampled ImageNet images for the same classes), CIFAR-100~\cite{krizhevsky2009cifar} (similar to CIFAR-10 but with 100 classes), and SVHN~\cite{netzer2011svhn} (consisting of 99,289 color images of house numbers from the Google Street View dataset). Additionally, we also choose two text datasets, which are used for training classification models and testing attacks, including Location~\cite{yang2015nationtelescope, yang2016participatory} (containing location ``check-in" records of mobile users in the Foursquare social network) and Texas~\cite{texas} (presented in the Hospital Discharge Data Public Use Data File provided by the Texas Department of State Health Services). All datasets are divided into three equal-sized parts: \(\mathcal{D}_\text{target}\), \(\mathcal{D}_\text{shadow}\), and \(\mathcal{D}_\text{reference}\). We have observed that some previous work may overlook the significant impact of the size of the attack dataset on the attack performance~\cite{liu2022membership}. Generally, a larger attack dataset leads to a better approximation of sample difficulty. Therefore, when evaluating various attacks, it is crucial to ensure that different attack methods have seen an equal number of samples during the training process.

\paragraph{Network Architecture.} For image datasets, we consider four commons architectures: VGG16~\cite{simonyan2014veryvgg16}, ResNet50~\cite{he2016deepresnet}, DenseNet121~\cite{huang2017densely}, and MobileNetV2~\cite{sandler2018mobilenetv2}. For text datasets, we train a model with two fully connected layers for classification. We use the SGD algorithm to train the models, with a learning rate (lr) set to 0.1, momentum set to 0.9, and weight decay~\cite{krogh1991simple} set to 5e-4. We also apply a cosine learning rate schedule~\cite{loshchilov2016sgdr} for optimization. Data augmentation~\cite{cubuk2018autoaugment} is enabled during the training of the target models to enhance their generalization. For the scoring model, we train a 4-layer
MLP with a single output channel.

\begin{figure*}[ht]
    \centering
    \subfloat[CIFAR-10]{\includegraphics[width=1.7in]{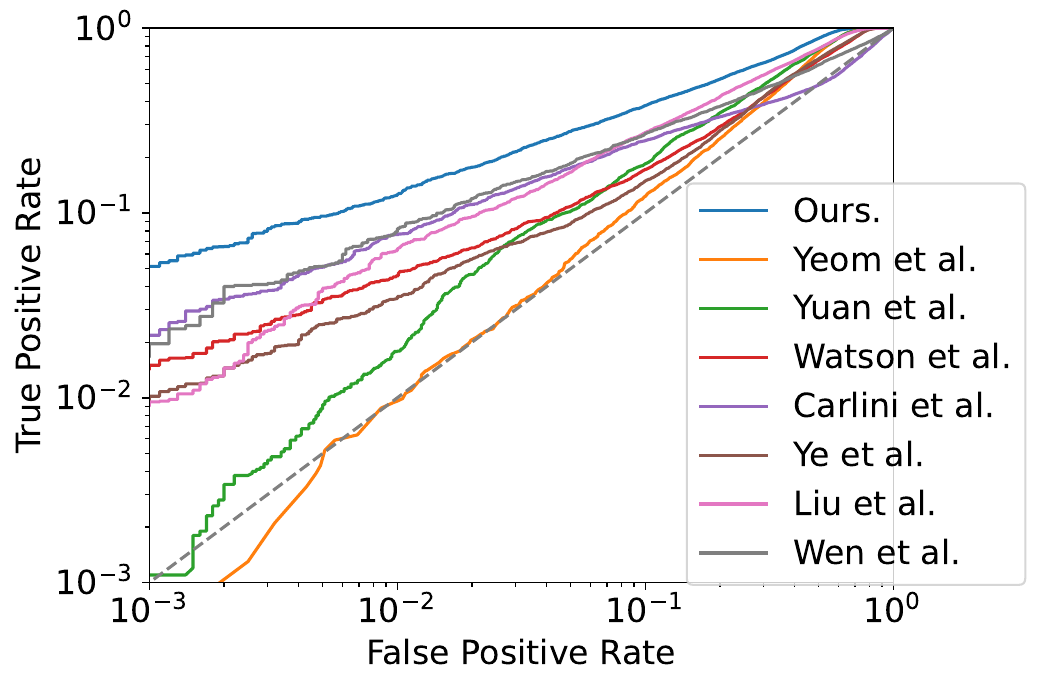} }
    \subfloat[CIFAR-100]{\includegraphics[width=1.7in]{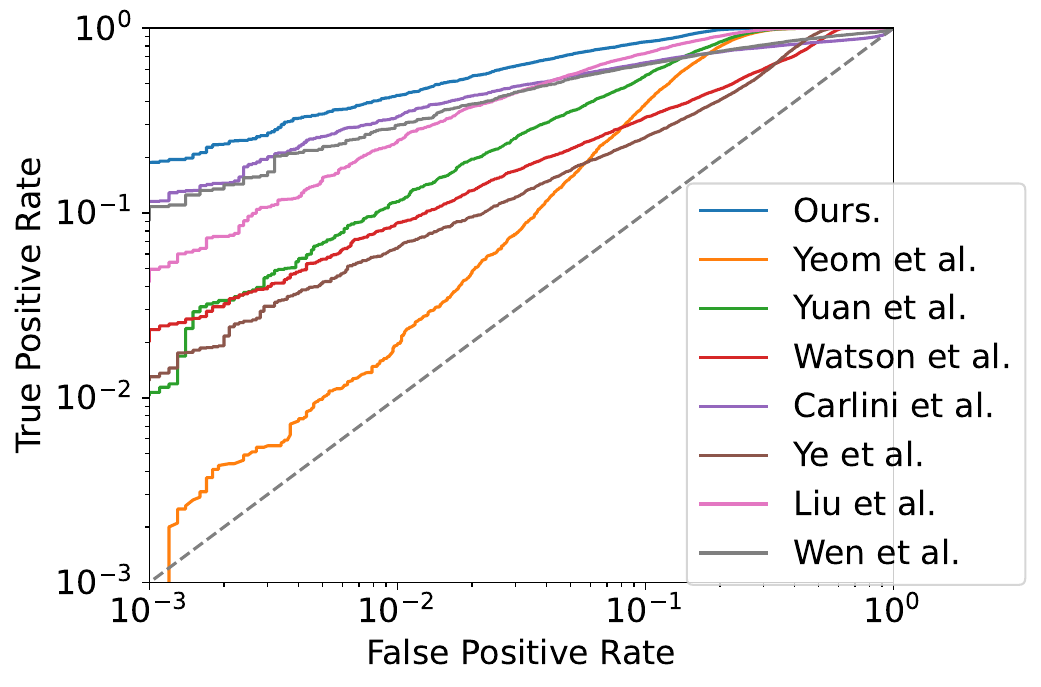} }
    \subfloat[CINIC-10]{\includegraphics[width=1.7in]{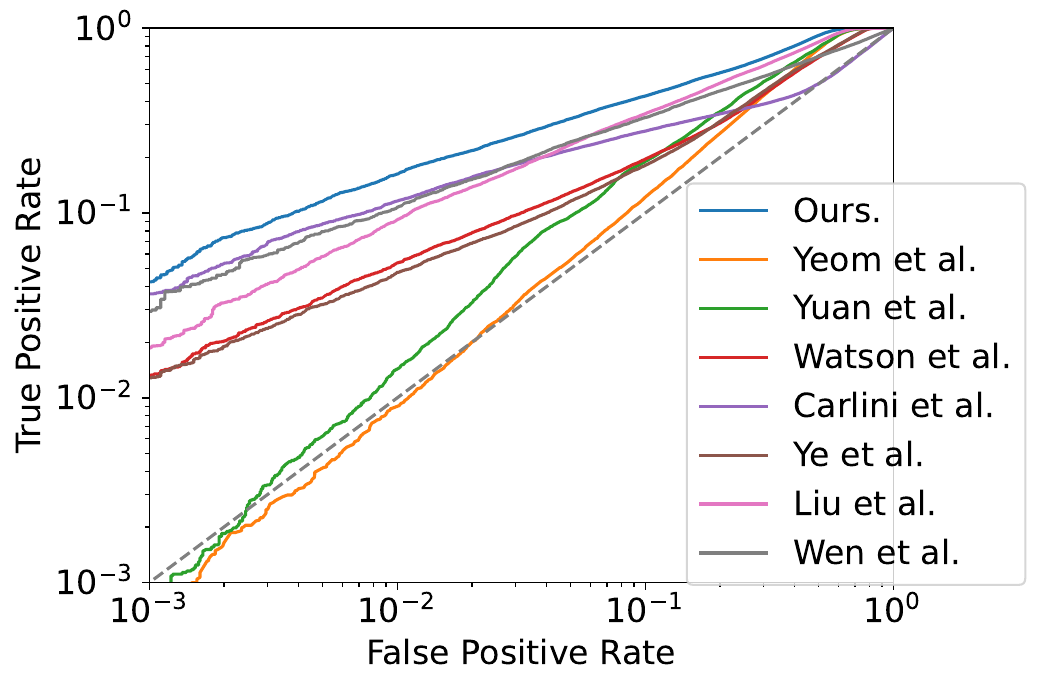} }
    \subfloat[SVHN]{\includegraphics[width=1.7in]{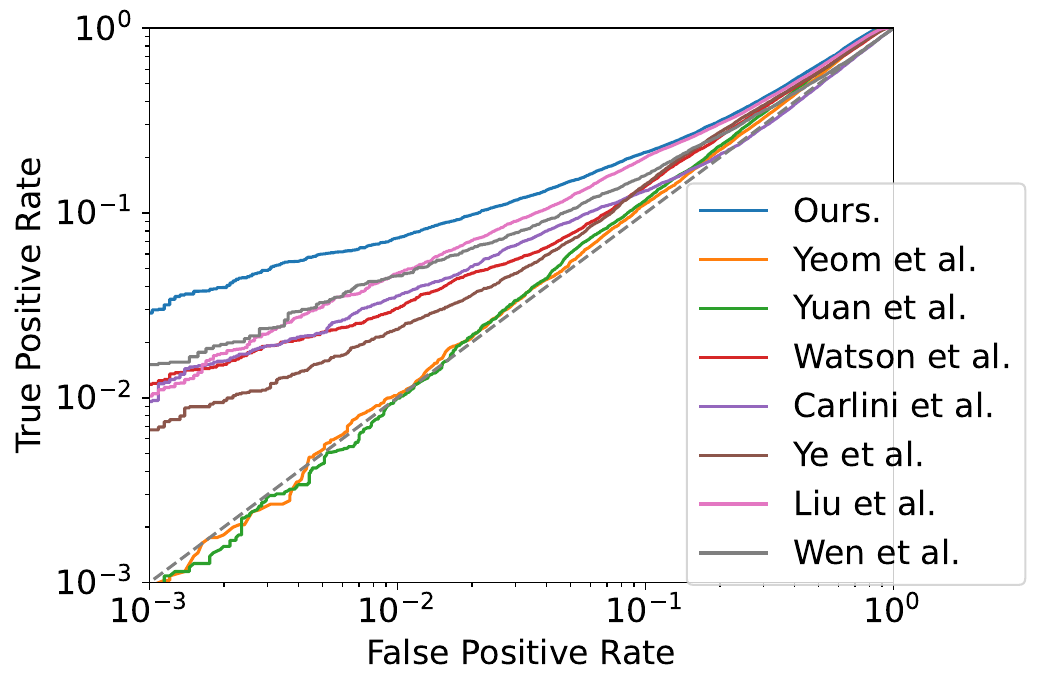} }
    \caption{The ROC curves of attack results on VGG16 models trained on four benchmark datasets.}
\label{Figure: ROC Curves of VGG16}
\end{figure*}
\begin{table*}[!t]
\centering
\setlength{\tabcolsep}{4.0pt}
\caption{The attack performances of different attacks on VGG16 models trained on four benchmark datasets. Additional attack results for other model architectures can be found in Appendix.\ref{Additional Experimental Results on Other Models}.}
\scalebox{0.87}
{
\begin{tabular}{l|cccccccccccc}
\toprule
Attack & \multicolumn{4}{c}{TPR @ 0.1\% FPR}&  \multicolumn{4}{c}{AUC}& \multicolumn{4}{c}{Balanced Accuracy}\\
\cmidrule(l{5pt}r{5pt}){2-5}\cmidrule(l{5pt}r{5pt}){6-9}\cmidrule(l{5pt}r{0pt}){10-13}
method& CIFAR-10& CIFAR-100& CINIC-10& SVHN&CIFAR-10& CIFAR-100& CINIC-10& SVHN&CIFAR-10& CIFAR-100& CINIC-10& SVHN\\
\midrule
Yeom et al.~\cite{yeom2018privacy}&0.0\%&0.1\%&0.1\%&0.1\%&0.643&0.866&0.660&0.552&64.1\%&83.4\%&65.8\%&55.8\%\\
Yuan et al.~\cite{yuan2022membership}&0.1\%&1.0\%&0.1\%&0.1\%&0.680&0.895&0.691&0.562&64.6\%&84.0\%&66.1\%&55.7\%\\
Watson et al.~\cite{watson2021importance}&1.4\%&2.0\%&1.3\%&1.2\%&0.629&0.750&0.645&0.566&58.6\%&68.9\%&58.9\%&53.4\%\\
Carlini et al.~\cite{carlini2022membership}&2.2\%&11.5\%&3.6\%&1.0\%&0.534&0.807&0.547&0.500&57.5\%&78.0\%&59.0\%&52.1\%\\
Ye et al.~\cite{ye2022enhanced}&1.0\%&1.2\%&1.3\%&0.5\%&0.629&0.752&0.649&0.571&59.2\%&71.9\%&59.3\%&53.3\%\\
Liu et al.~\cite{liu2022membership}&0.9\%&4.2\%&1.8\%&1.0\%&0.708&0.929&0.755&0.600&64.2\%&85.4\%&67.2\%&56.0\%\\
Wen et al.~\cite{wen2022canary}&1.7\%&9.1\%&2.9\%&1.5\%&0.610&0.837&0.665&0.532&59.0\%&77.6\%&62.9\%&53.5\%\\
\midrule
Ours&\textbf{5.1\%}&\textbf{18.8\%}&\textbf{4.9\%}&\textbf{2.9\%}&\textbf{0.776}&\textbf{0.958}&\textbf{0.799}&\textbf{0.618}&\textbf{69.1\%}&\textbf{89.1\%}&\textbf{70.5\%}&\textbf{57.1\%}\\
\bottomrule
\end{tabular}
}
\label{table:vgg16 attack performances}
\end{table*}

\paragraph{Attack Baselines.} We compare our RAPID with seven state-of-the-art or representative attack methods. Among them, Yeom et al.~\cite{yeom2018privacy} leverage the loss of the target model for decision boundary estimation. Yuan et al.~\cite{yuan2022membership} propose a new signal called sensitivity, which exhibits a larger gap between members and non-members. Both Watson et al.~\cite{watson2021importance} and Ye et al.~\cite{ye2022enhanced} employ difficulty calibration. The latter emphasizes the reliance of membership scores on the target model and uses models distilled from the target model as reference models. Liu et al.~\cite{liu2022membership} take the first step to exploit the information from the training trajectory to conduct membership inference attacks and achieve advanced performance. Both the methods proposed by Ye et al. and Liu et al. rely on the property that the self-distilled reference model is similar to the target model so that for non-members \(\mu_\text{tar}\) and \(\mu_\text{ref}\) will be closer. We also compare RAPID with LiRA proposed by Carlini et al.~\cite{carlini2022membership} and Canary proposed by Wen et al.~\cite{wen2022canary}, both using Gaussian likelihood
estimate and currently achieving the best performance. Specifically, LiRA utilizes augmentations of the target sample to compute statistics, while Canary uses adversarial tools to directly optimize for queries that are discriminative.

\begin{table*}[!t]
\newcommand{\tabincell}[2]{\begin{tabular}{@{}#1@{}}#2\end{tabular}}
\centering
\setlength{\tabcolsep}{2.5pt}
\caption{Time cost of all attacks against a VGG16 model trained on CIFAR-10.}
\scalebox{1.05}
{
\begin{tabular}{l|cccccccccc}
\toprule
Attack Method&~\cite{yeom2018privacy}&~\cite{yuan2022membership}&~\cite{watson2021importance}&LiRA offline~\cite{carlini2022membership}&LiRA online~\cite{carlini2022membership}&~\cite{ye2022enhanced}&~\cite{liu2022membership}&Canary offline~\cite{wen2022canary}&Canary online~\cite{wen2022canary}&\textbf{ours.}\\
\midrule
Time Cost/h& 0.22& 0.47& 0.46& 13.82& $>$200000& 0.87& 0.51& 38.5& $>$100000&\textbf{0.58}\\
\bottomrule
\end{tabular}
}
\label{table:time cost analysis}
\end{table*}

\paragraph{Attack Setup.} In the main experiments, we train 4 reference models on \(\mathcal{D}_\text{reference}\), each with a different random initialization. We compute the average of the membership scores obtained from these reference models to calculate the calibrated membership scores. For all models, we utilize the multiple queries technique, where the average membership scores obtained from all queries represent per model's output scores. To ensure fairness in comparison, we re-implement existing attacks (except LiRA and Canary) using the same number of reference models as ours, if they have claimed that their performance is related to the number of reference models in the papers. When evaluating LiRA and Canary, we follow the original papers' setting for the number of reference models (128 and 64 respectively). We also ensure that LiRA queries the same number of times as RAPID (8 times). Since optimizing hyperparameters significantly affects the performance of Canary, we follow its original settings.

\paragraph{Why Not Online Version?} In practice, both LiRA and Canary have an offline version and an online version. For instance, LiRA online firstly trains 256 reference models, half of which are \textit{IN models} trained on datasets that include the target record, and the other half are \textit{OUT models} trained on datasets that do not include the target record. Then, LiRA fits two Gaussian distributions to the confidences of the IN and OUT models on the target record. Finally, it queries the confidence of \(\mathcal{M}_\text{target}\) on the target record and outputs a likelihood-ratio test. However, LiRA offline only trains 128 OUT models and outputs a one-sided hypothesis test. When evaluating LiRA and Canary in our main experiments, we implement the offline version of them. This is because LiRA (Canary) online requires training 128 (64) IN models for each target sample, which is not so feasible for common attackers. More discussions on why comparing with LiRA (Canary) offline is reasonable can be found in Attack Cost Analysis of Section~\ref{sec: 5.2}.

\subsection{Experimental Results}
\label{sec: 5.2}
Finally, we present the performance of our attack in the black-box scenario, comparing it to the seven advanced baselines~\cite{yeom2018privacy, watson2021importance, yuan2022membership, carlini2022membership, ye2022enhanced, liu2022membership, wen2022canary}. Furthermore, we provide a detailed attack cost analysis of all attacks. Lastly, we provide the results of all attacks against models using DP-SGD~\cite{abadi2016deep}. Table \ref{table:target model performances} reports the accuracy of \(\mathcal{M}_\text{target}\).

\paragraph{Main Evaluation.} Compared to the latest representative works, our proposed attack outperforms. Figure \ref{Figure: ROC Curves of VGG16} demonstrates the superior performance of our attack in the low FPR regime. This holds even versus the attacks using Gaussian likelihood estimate~\cite{carlini2022membership,wen2022canary}, which require training a large number of reference models. Table \ref{table:vgg16 attack performances} presents the same advanced performance of our attack in terms of average metrics, surpassing previous attacks~\cite{yeom2018privacy, watson2021importance, yuan2022membership, carlini2022membership, ye2022enhanced, liu2022membership, wen2022canary} by a significant margin. For example, over CIFAR-100 RAPID elevates the best TPR @0.1\% FPR from 11.5\% to 18.8\%, best AUC from 0.929 to 0.958, and best Acc from 85.4\% to 89.1\%. We posit that this represents a significant advancement in the MIA domain, as previous work has struggled to achieve optimal performance across all metrics simultaneously. For instance, while LiRA and Canary offline achieve significant breakthroughs in currently recommended TPR at 0.1\% FPR, they even fall short of the initial loss attack~\cite{yeom2018privacy} in metrics reflecting average-case success. Additional attack results for other model architectures and datasets can be found in Appendix.\ref{Additional Experimental Results on Other Models} and Appendix.\ref{Additional Experimental Results on Location and Texas Datasets}. 

\paragraph{Attack Cost Analysis.} To shed light on the practicality of existing state-of-the-art work~\cite{carlini2022membership, ye2022enhanced, liu2022membership, wen2022canary}
and our proposed attack, we will provide an analysis in two aspects: query cost and computational cost. In distilled-based attacks~\cite{ye2022enhanced, liu2022membership}, assuming the distillation dataset size is \(N\) and the number of distillation rounds is \(E\), if the attack targets \(n\) records, the attack would require \(NE+n\) queries on the target model. In contrast, our attack only requires \(8n\) queries on the target model. Specifically, taking CINIC-10 as an example, the cost of our attack for 20000 sample points reduces to approximately 1/42. As for computational cost, our proposed RAPID only requires training 4 reference models and a shadow model to achieve better performance compared to LiRA~\cite{carlini2022membership}, which requires training at least 128 models. This reduces the computational cost to approximately 1/25 (and potentially lower). To provide an explicit time complexity analysis for all attacks, we report the total time cost of various attacks against a VGG16 model trained on CIFAR-10 using a single NVIDIA GeForce RTX 3070 Ti in Table \ref{table:time cost analysis}. The time cost of LiRA (Canary) online is approximately proportional to the number of samples attacked. Thus, its time cost is calculated theoretically by measuring the time required to attack one sample. The astronomical computational overhead of LiRA (Canary) online renders it an infeasible attack---the adversary needs to train 128 IN models for each potential member at inference time. Therefore, we use LiRA (Canary) offline as the state-of-the-art baselines in our main experiments. 

In practice, Carlini et al. \cite{carlini2022membership} use a clever method that circumvents the necessity to train 128 models for each point to evaluate the theoretical performance of LiRA online. However, we note that the implementation in their code repository\footnote{\url{https://github.com/tensorflow/privacy/tree/master/research/mi_lira_2021}} 1) relaxes the assumption that the attack set contains only non-members, and (2) potentially makes LiRA online advantaged more (than traditional implementations) as the IN/OUT models are highly similar to the target model. To ensure fairness in comparison, we have placed a detailed discussion of the theoretical performance gap between RAPID and LiRA online in Section \ref{sec: 8}.

\begin{table}[!t]
\newcommand{\tabincell}[2]{\begin{tabular}{@{}#1@{}}#2\end{tabular}}
\centering
\setlength{\tabcolsep}{2.5pt}
\caption{Attack performance of RAPID against a DenseNet121 model trained on CIFAR-10 using DP-SGD.}
\scalebox{0.8}
{
\begin{tabular}{l|ccccc}
\toprule
Noise& \multicolumn{5}{c}{C = 10}\\
Multiplier ($\sigma$)& $\epsilon$& Model Acc& Attack Acc& TPR @ 0.1\% FPR& Attack AUC\\
\midrule
0.0& $\infty$&77.1\%& 67.7\%& 3.0\%& 0.756\\
0.1& >5000&66.8\%& 54.0\%& 0.2\%& 0.552\\
0.2& >1000&58.6\%& 51.2\%& 0.2\%& 0.516\\
0.5& >100 &44.9\%& 50.4\%& 0.2\%& 0.507\\
1.0& 8    &30.2\%& 49.9\%& 0.1\%& 0.502\\
\bottomrule
\end{tabular}
}
\label{table:our attack against DPSGD}
\end{table}

\begin{figure}[t]
    \centering
    \includegraphics[width=0.48\textwidth]{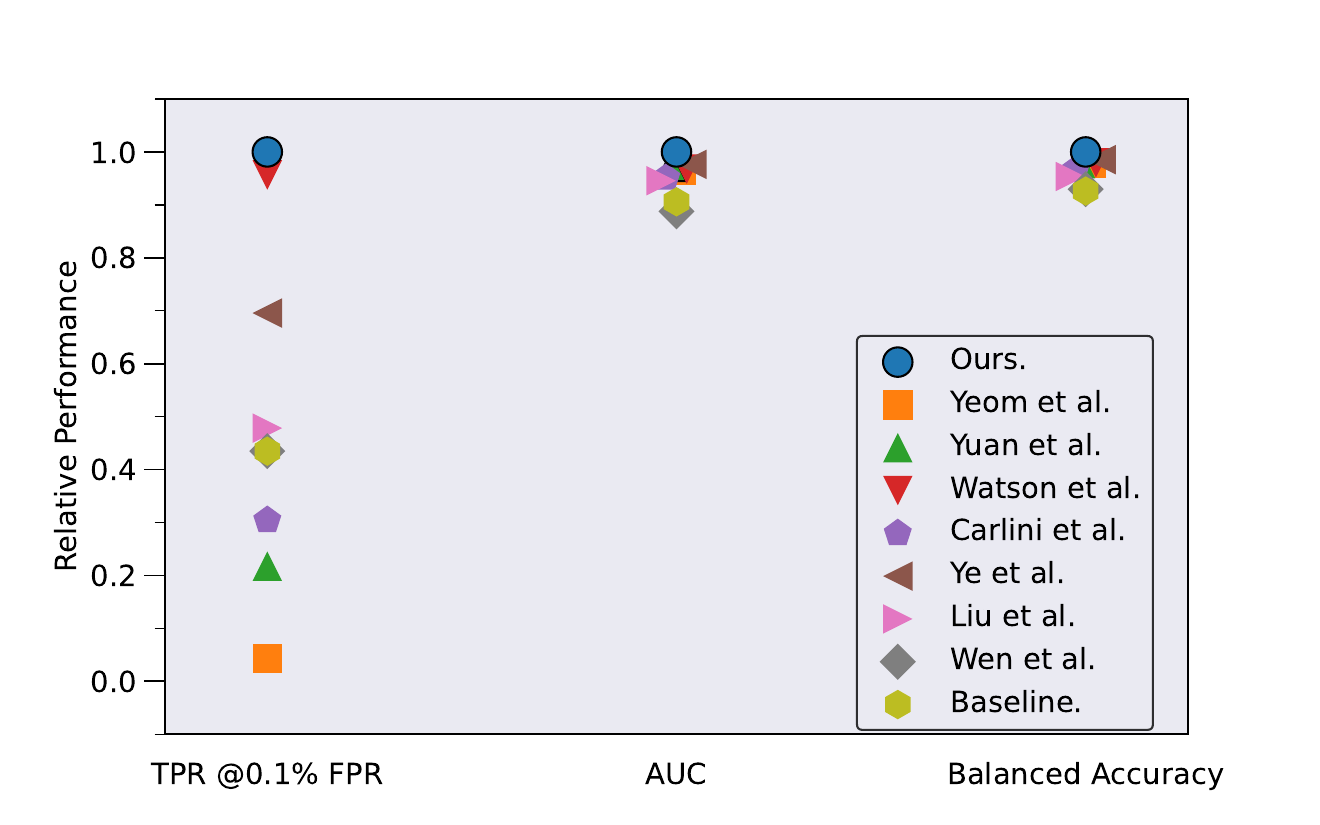}
    \caption{Attack performance of prior works and our attack against a DenseNet121 model trained on CIFAR-10 using DP-SGD. The noise multiplier \(\sigma\) is set to 0.1. Additional attack results for other \(\sigma\) can be found in Appendix.\ref{sec:Attacking DP-SGD}.}
    \label{Figure: DPSGD}
\end{figure}

\paragraph{Attack Against DP-SGD} Differential privacy~\cite{dwork2006calibrating} is a widely used defense mechanism against all privacy leakage attacks~\cite{song2019auditing, lecuyer2019certified, krishna2019thieves}. It imposes theoretical bounds on the success rate of MIAs by directly restricting the ability to distinguish between two neighboring datasets (differing only in the inclusion or exclusion of a particular sample). This is directly related to MIAs. Previous studies have also explored this scenario~\cite{liu2022membership, carlini2022membership}, and we follow their investigation to examine the defensive effect of the DP-SGD training algorithm~\cite{abadi2016deep} on our attack. We fixed the clipping norm to 10 and evaluated the performance of prior works and our attack on a DenseNet121 model trained on the CIFAR-10 dataset. The privacy budget \(\epsilon\) can be controlled by varying the noise multiplier parameter. From Table \ref{table:our attack against DPSGD} and Figure \ref{Figure: DPSGD}, we can observe that DP-SGD indeed effectively defends against all MIAs. However, DP-SGD significantly reduces the classification accuracy of the target model under high clipping norms, even when the noise multiplier is set to 0.1. We should thus carefully consider the trade-off between the defense level achieved by differential privacy and the loss of model accuracy. We primarily focus on the scenario where \(\sigma\) (noise multiplier) is set to 0.1 to evaluate the defense level of DP against existing attacks since this setting maintains an acceptable model accuracy. It can be observed that while the gap between different attacks has narrowed, our attack continues to outperform other works across all metrics. Our work presents a greater challenge to DP-SGD in the better trade-off between defense level and model performance.

\subsection{Attack Against LLMs}
\label{sec: 5.3}
In the realm of LLMs, MIAs can assess the degree of privacy leakage in both the pre-training and fine-tuning stages. Pre-training is primarily conducted on publicly available datasets, and the data used to train the model is often public knowledge. Fine-tuning typically occurs on smaller, and more private datasets. Therefore, our primary focus is on the fine-tuning phase, where full model fine-tuning and prompt-based learning~\cite{brown2020language, radford2019language} are two commonly used methods. Previous work has already pointed out that the privacy risk of prompted models exceeds that of fine-tuned models at the same utility levels~\cite{duan2023privacy}, and we are thus interested in whether our proposed RAPID can launch an effective attack against fine-tuned LLMs or not.

\paragraph{Setup.} We fine-tune BERT~\cite{devlin2018BERT} to solve three standard downstream text classification tasks: cola~\cite{wang2018glue}, cb~\cite{de2019commitmentbank}, and mrpc~\cite{wang2018glue}. This is because BERT has demonstrated strong generalization capabilities on these classification tasks. To investigate the impact of model size on membership risk, we conduct attack evaluations on both the BERT-base version (total Parameters=110M) and the BERT-large version (total Parameters=340M). Within all our experiments, the learning rate (lr) is set to 3e-5 and weight decay~\cite{krogh1991simple} is set to 5e-4 in the training process. We fine-tune the model for 20 epochs and use the checkpoint with the highest validation accuracy during tuning. We report the fine-tuning results in Table \ref{table:LLMs performances}. In the attack setup, we follow the data splitting method outlined in Section~\ref{sec: 5.1} and only fine-tune two reference models on \(\mathcal{D}_\text{reference}\) for our attack. The technique of multiple queries is not employed because there is no natural data augmentation available in the text domain as there is in the image domain. However, Mattern et al. ~\cite{mattern2023membership} have recently proposed a neighborhood attack that uses synthetically generated neighboring texts. This aligns closely with our idea, implying RAPID's potential for further enhancement in attacking LLMs.

\paragraph{Experimental Results.} We compare our attack to the original loss-based attack in~\cite{duan2023privacy} and attacks with difficulty calibration in~\cite{watson2021importance, mattern2023membership} as other baselines do not take this scenario into consideration. The results in Table \ref{table:attack on mrpc} demonstrate that RAPID still outperforms other baselines in attacking well-fine-tuned LLMs. However, the advantage of our attack is observably reduced compared to that of the computer vision domain, especially in terms of TPR at low FPR. One possible reason is that LLMs, due to their strong generalization capabilities obtained from the pre-training phase, result in small prediction losses for most non-members. In other words, the number of misclassified non-members (due to difficulty calibration) that can be directly corrected using the original membership scores is smaller. This is consistent with the worse TPR results for BERT-large compared to BERT-base, as BERT-large has a larger model capacity and stronger generalization abilities. Note that Carlini et al.~\cite{carlini2021extracting} have demonstrated that larger pre-trained language models would memorize more training data, which contrasts with the experimental results in Table \ref{table:attack on mrpc}. We speculate that this is because the memorization principles of LLMs differ during the pre-training and fine-tuning stages. We have also observed that even with a larger training-testing accuracy gap compared to models trained on SVHN (see Table~\ref{table:target model performances}), the TPRs of all attacks against LLMs become generally worse, which contradicts traditional views. We hypothesize that the target dataset itself probably has a quite small proportion of outliers (hard samples), making the distribution of outputs for member points similar between the target models and reference models. We argue that \textbf{the inherent distribution properties of the dataset also significantly influence the attack's TPR at a given FPR, not only the level of overfitting.} For more experimental results, please refer to Appendix.\ref{Additional Experimental Results on cola and cb}.

\begin{table}[t]
\newcommand{\tabincell}[2]{\begin{tabular}{@{}#1@{}}#2\end{tabular}}
\centering
\setlength{\tabcolsep}{2.5pt}
\caption{The classification accuracy of BERT-base and BERT-large fine-tuned on different datasets.}
\scalebox{0.89}
{
\begin{tabular}{l|cccccc}
\toprule
Dataset&  \multicolumn{2}{c}{cola}&  \multicolumn{2}{c}{cb}& \multicolumn{2}{c}{mrpc}\\
\cmidrule(l{5pt}r{5pt}){2-3}\cmidrule(l{5pt}r{5pt}){4-5}\cmidrule(l{5pt}r{0pt}){6-7}
Model&Train acc&Test acc&Train acc&Test acc&Train acc&Test acc\\
\midrule
BERT-base&  {98.0\%}& 80.4\%&  {99.8\%}& 79.2\%&  {99.5\%}& 77.8\%\\
BERT-large&  {100.0\%}& 86.3\%&  {99.9\%}& 82.6\%&  {99.7\%}& 81.4\%\\
\bottomrule
\end{tabular}
}
\label{table:LLMs performances}
\end{table}

\begin{table}[!t]
\centering
\setlength{\tabcolsep}{2.5pt}
\caption{The attack results of BERT-base and BERT-large fine-tuned on mrpc.}
\scalebox{0.68}
{
\begin{tabular}{l|cccccc}
\toprule
 & \multicolumn{2}{c}{TPR @ 0.1\% FPR}&  \multicolumn{2}{c}{AUC}& \multicolumn{2}{c}{Balanced Accuracy}\\
\cmidrule(l{5pt}r{5pt}){2-3}\cmidrule(l{5pt}r{5pt}){4-5}\cmidrule(l{5pt}r{0pt}){6-7}
Attack Method& BERT-base& BERT-large& BERT-base& BERT-large& BERT-base& BERT-large\\
\midrule
Duan et al.~\cite{duan2023privacy}&0.2\%&0.1\%&0.686&0.689&63.1\%&59.9\%\\
Watson et al.~\cite{watson2021importance}&0.4\%&\textbf{0.2}\%&0.654&0.654&59.0\%&58.4\%\\
\midrule
Ours&\textbf{1.1\%}&\textbf{0.2\%}&\textbf{0.745}&\textbf{0.700}&\textbf{66.7\%}&\textbf{60.1\%}\\
\bottomrule
\end{tabular}
}
\label{table:attack on mrpc}
\end{table}

\section{Ablation Study}
\label{sec: 6}
In this section, we conduct extensive experiments to investigate the specific impact of each component on the final performance. We aim to further substantiate our explanation in Section \ref{sec: 3} regarding the suboptimality of difficulty calibration. Specifically, we start by exploring the impact of reference models and the signal function employed. Then we discuss the effects of random sampling and the number of \(\mathcal{M}_\text{reference}\) on the attack. We also examine the impact of varying numbers of queries. Lastly, we attempt to relax two common assumptions regarding the same architecture to \(\mathcal{M}_\text{target}\) and i.i.d. \(\mathcal{D}_\text{attack}\) used by the attacker to demonstrate the efficacy of our attack in more realistic scenarios. In our ablation studies, we utilize the CINIC-10 dataset by default unless otherwise stated.

\begin{table*}[!t]
\centering
\setlength{\tabcolsep}{4.0pt}
\caption{Comparison of the performance using prior calibrated membership scores and ours on a VGG16 model trained on CINIC-10. We evaluate different signal functions and reference models.}
\scalebox{0.85}
{
\begin{tabular}{l|cccccccccccc}
\toprule
 & \multicolumn{4}{c}{Loss}&  \multicolumn{4}{c}{Conf}& \multicolumn{4}{c}{GN}\\
\cmidrule(l{5pt}r{5pt}){2-5}\cmidrule(l{5pt}r{5pt}){6-9}\cmidrule(l{5pt}r{0pt}){10-13}
Reference& Calibrated& Our& Calibrated& Our&Calibrated& Our& Calibrated& Our&Calibrated& Our& Calibrated& Our\\
Model& Acc& Acc& AUC& AUC&Acc& Acc& AUC& AUC&Acc& Acc& AUC& AUC\\
\midrule
Trivial Model~\cite{watson2021importance}&59.6\%&\textbf{66.6\%}&0.658&\textbf{0.739}&60.6\%&\textbf{62.6\%}&0.636&\textbf{0.687}&61.9\%&\textbf{66.9\%}&0.662&\textbf{0.746}\\
\midrule
Distilled Model~\cite{ye2022enhanced}&59.4\%&\textbf{66.4\%}&0.654&\textbf{0.749}&56.8\%&\textbf{62.1\%}&0.604&\textbf{0.702}&63.3\%&\textbf{67.1\%}&0.680&\textbf{0.756}\\
\bottomrule
\end{tabular}
}
\label{table:general enhancement}
\end{table*}

\subsection{Reference Model and Signal Function}
\begin{figure}[t]
    \centering
    \subfloat[Trivial Model]{\includegraphics[width=1.7in]{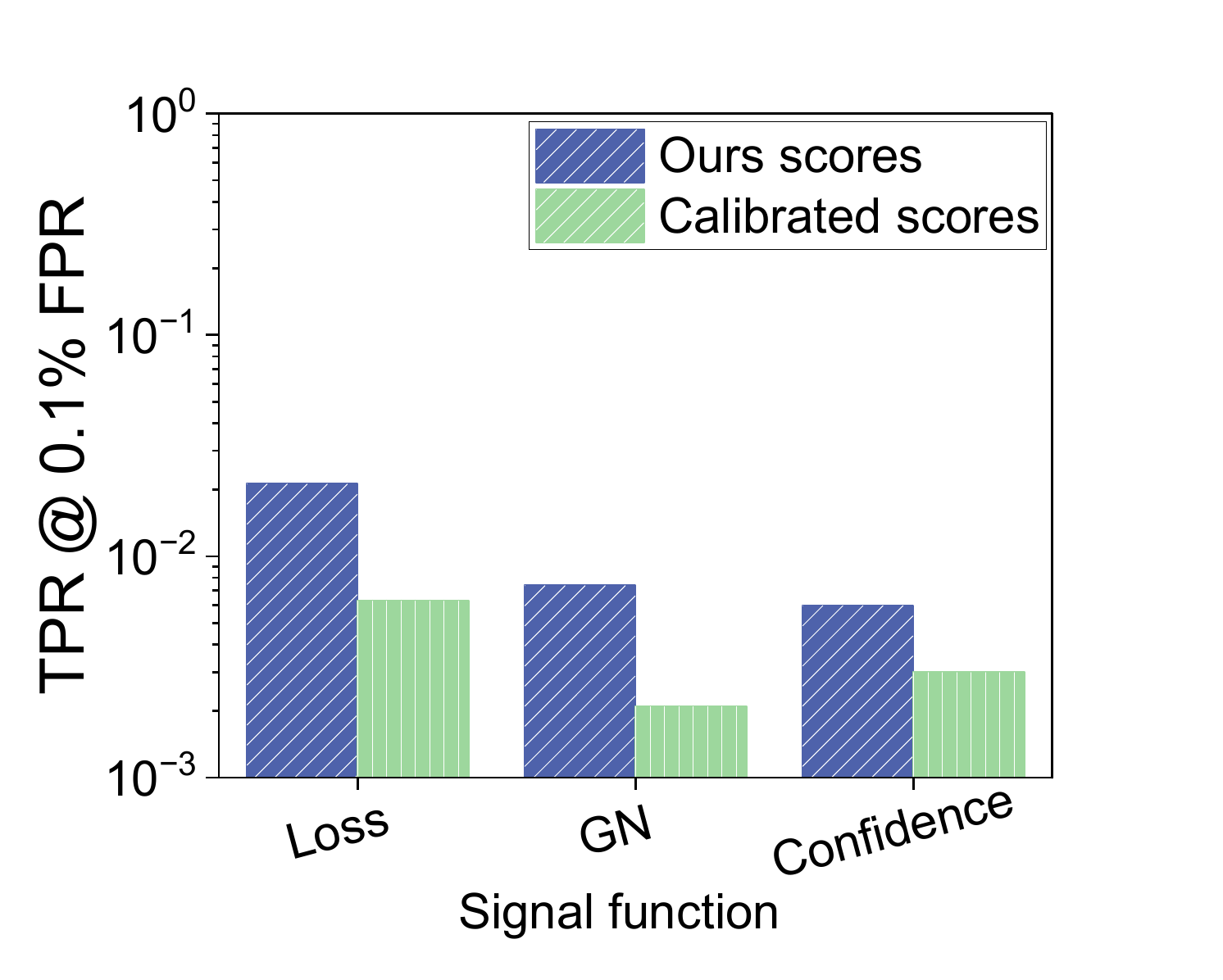} }
    \subfloat[Distilled Model]{\includegraphics[width=1.7in]{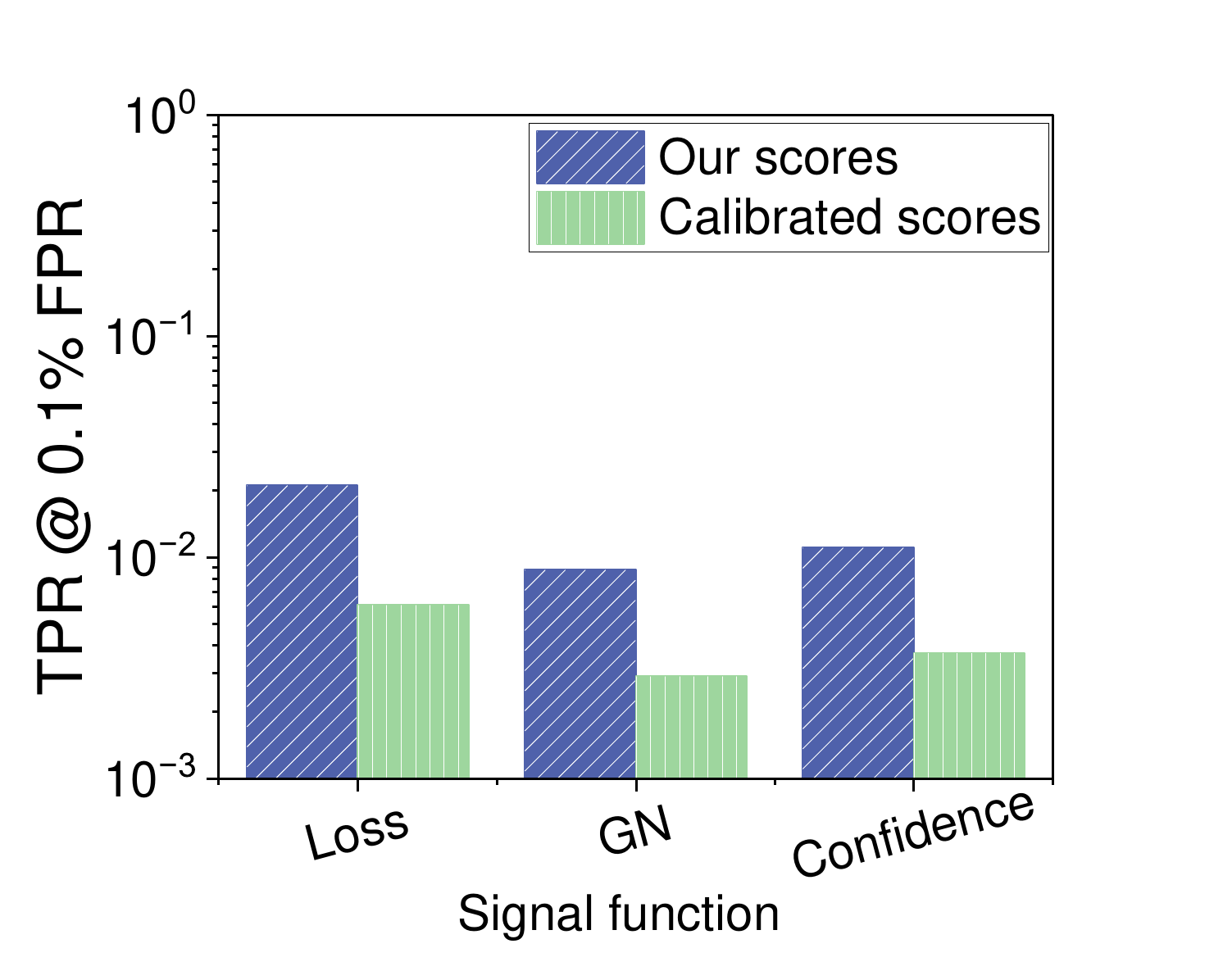} }
    \caption{Our proposed method significantly improves the TPR at low FPR compared to solely using calibrated membership scores.}
    \label{Figure 5}
\end{figure}

In practice, the adversary can select different reference models and signal functions for difficulty calibration. Common reference models include models trained from scratch on \(\mathcal{D}_\text{reference}\) (i.e., trivial models~\cite{watson2021importance}) and models distilled from the target model (i.e., distilled models~\cite{ye2022enhanced}). As for signal functions, common options include loss~\cite{yeom2018privacy}, confidence~\cite{salem2018ml}, and gradnorm~\cite{nasr2019comprehensive}. To investigate whether intrinsic errors in difficulty calibration are a pervasive phenomenon, we compare the different performances of attacks using only calibrated scores and attacks re-leveraging original scores across various reference models and signal function settings. Note that to emphasize the direct impact of the shortcut introduced by us, other enhancement techniques such as random sampling and multiple queries are not utilized. Table \ref{table:general enhancement} and Figure \ref{Figure 5} demonstrate that introducing a shortcut of \(S_{target}(x,y)\) effectively enhances the performance across all evaluation metrics. This justifies our claim that difficulty calibration represents a suboptimal approach and that original membership scores can directly correct errors it generates.
 
\begin{figure}[t]
    \centering
    \includegraphics[width=0.48\textwidth]{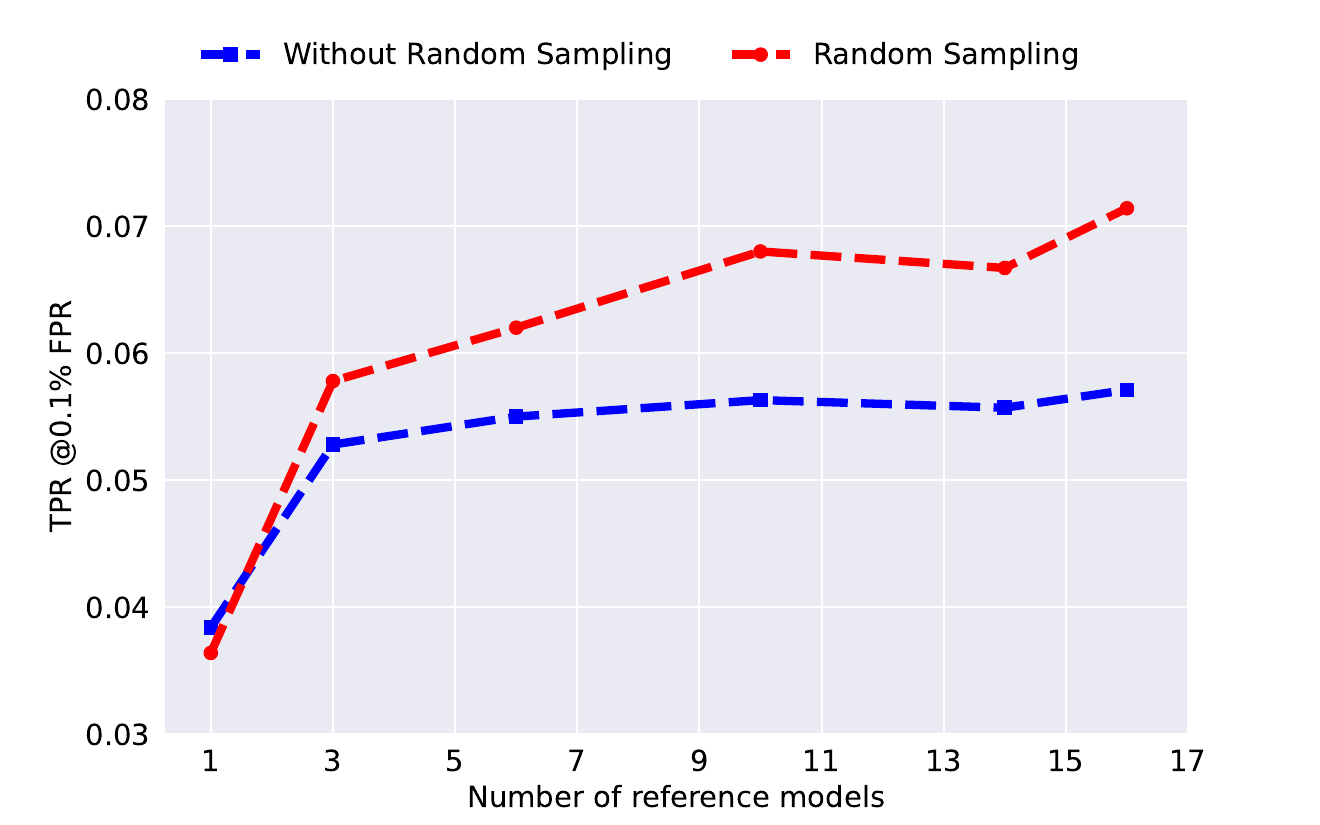}
    \caption{The attack performance exhibits apparent enhancement as the number of reference models increases, with diminishing returns. When using random sampling, the attack results have a higher upper bound.}
    \label{Figure random_sampling}
\end{figure}
\subsection{Random Sampling}
\begin{figure}[t]
    \centering
    \includegraphics[width=0.48\textwidth]{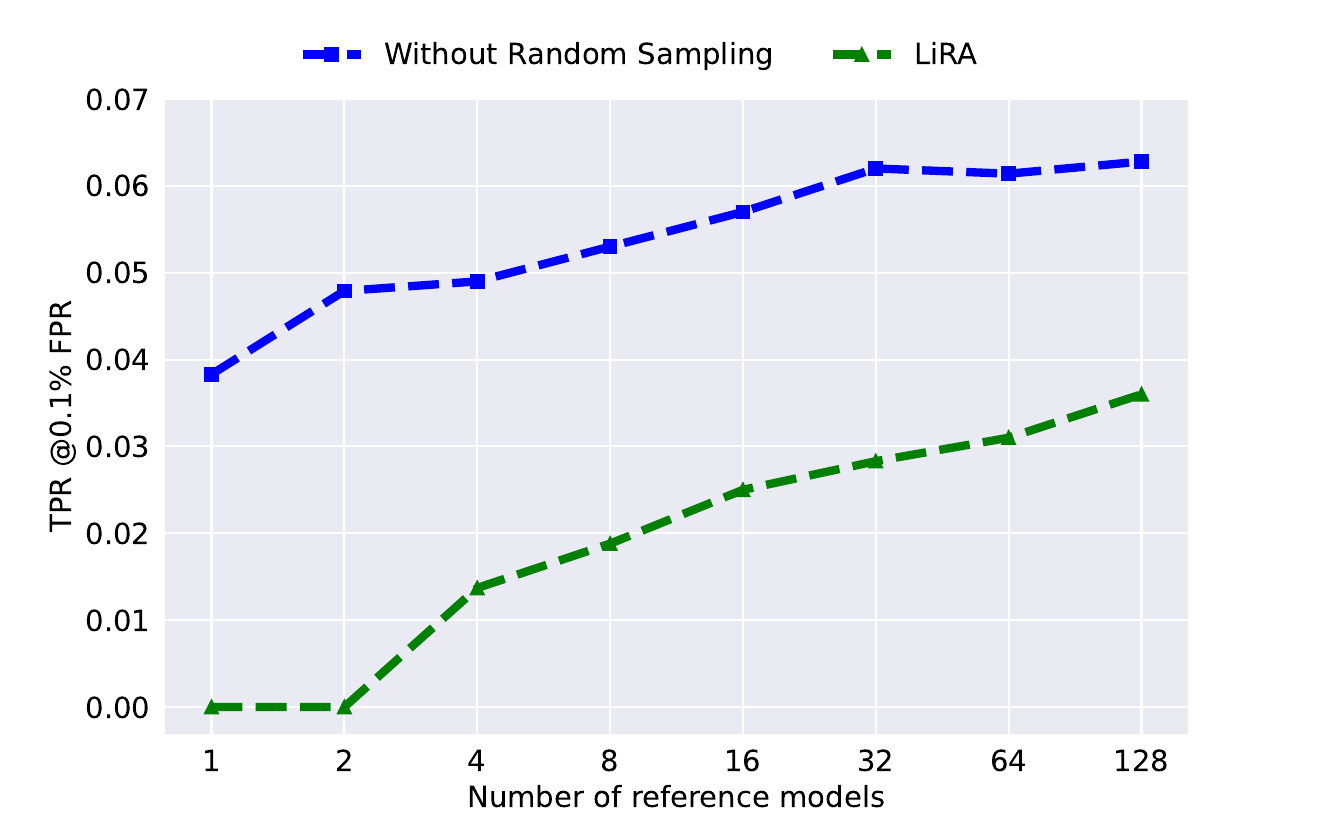}
    \caption{Both RAPID and LiRA exhibit enhanced attack performance as the number of reference models increases, with RAPID benefiting more from training additional reference models.}
    \label{Figure RAPIDvsLiRA}
\end{figure}

We have argued the adversary can significantly enhance the attack by training several \(\mathcal{M}_\text{reference}\) using random sampling when he has a larger \(\mathcal{D}_\text{attack}\) compared to \(\mathcal{D}_\text{target}\). In the worst case, the attacker can also achieve a slightly weaker improvement by changing the initialization parameters of these \(\mathcal{M}_\text{reference}\) trained on \(\mathcal{D}_\text{reference}\). We are interested in understanding the impact of the number of \(\mathcal{M}_\text{reference}\) on the attack results, both with and without random sampling. Figure \ref{Figure random_sampling} illustrates the TPR of our attack at a fixed FPR of 0.1\% as the number of \(\mathcal{M}_\text{reference}\) increases. As expected, a larger number of reference models leads to better attack performance. It is further enhanced when random sampling is employed, as having a larger number of seen data points when training \(\mathcal{M}_\text{reference}\) means the extent to the target point represented in \(\mathcal{D}_\text{reference}\), becomes more like that under the entire distribution \(\pi\). Training more than two reference models brings diminishing benefits as the averaged results gradually stabilize. Another question is whether the rate at which RAPID's attack success rate increases, relative to the associated attack cost, outpaces existing methods, and we use LiRA as the baseline to answer this question. Figure \ref{Figure RAPIDvsLiRA} demonstrates that RAPID benefits more from the ability to train increasing numbers of reference models. Specifically, LiRA requires training at least 32 reference models to capture the majority of the benefits.

\subsection{Multiple Queries}
\label{sec: 6.2}
\begin{figure}[t]
    \centering
    \includegraphics[width=0.48\textwidth]{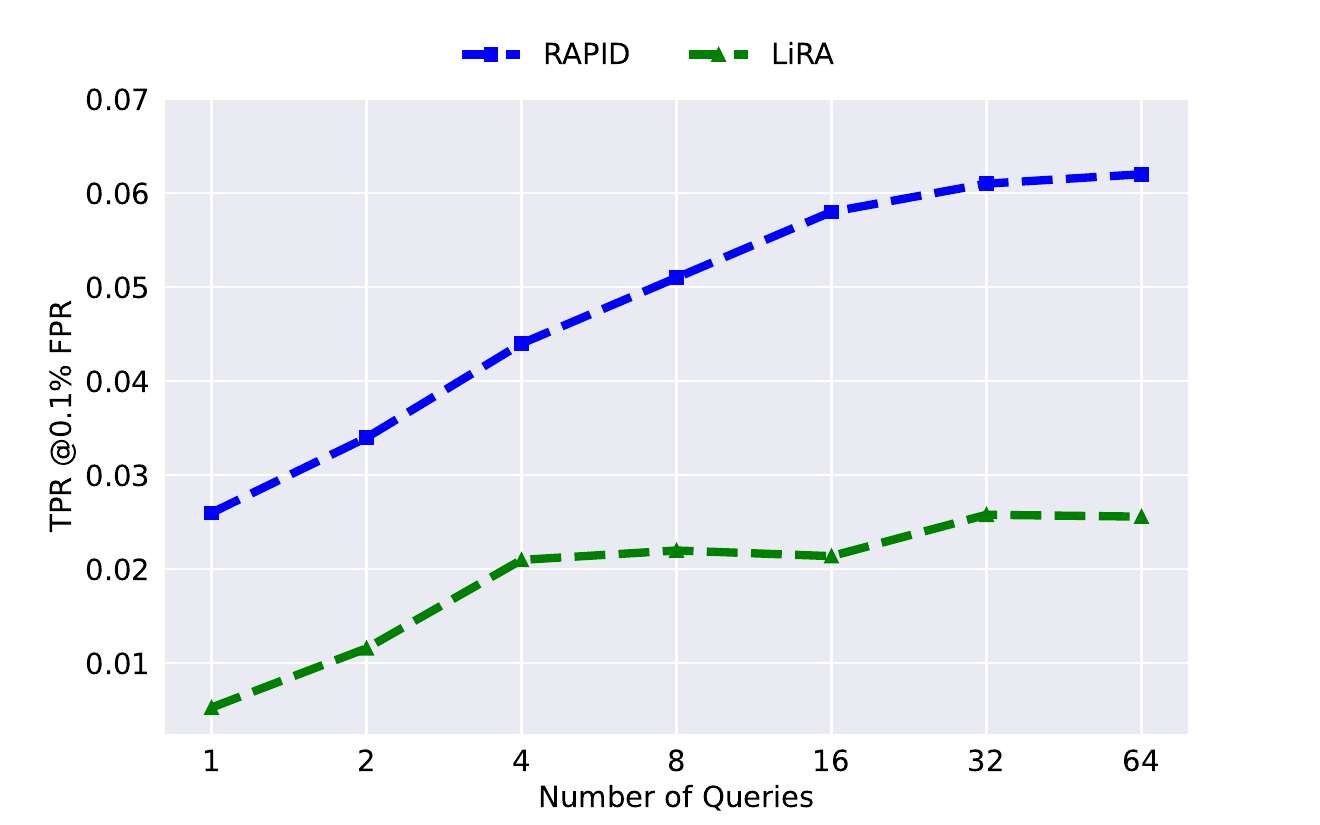}
    \caption{Multiple queries on the augmented versions of the target sample can significantly enhance the attack performances,  and RAPID consistently outperforms LiRA.}
    \label{Figure Multiple Queries}
\end{figure}

Previous work~\cite{carlini2022membership} has suggested that models are typically trained to minimize their loss on augmented versions of examples, which inspires the idea of conducting MIAs on augmented versions of examples that have been seen during training. However, the results of attacks against CIFAR-10 in Figure \ref{Figure Multiple Queries} are not entirely consistent with this explanation. Note that we average the membership scores obtained from multiple queries on the target model to obtain the final \(\mathcal{S}(x,y)\). The experimental results show that increasing the number of queries leads to diminishing improvements in attack performance. Under the previous explanation, the results of each query on different augmented versions of target samples should be independent and of equal importance, so that averaging results of multiple queries should not lead to such significant improvements. The final experimental results actually align perfectly with the analysis provided in Section \ref{sec: 4.3}. This is also why the trade-off observed with multiple queries is similar to that of random sampling. Furthermore, Figure \ref{Figure Multiple Queries} demonstrates that RAPID continues to outperform LiRA even with an increasing number of queries. This justifies our claim that averaging outputs is equally good for fitting multiple-dimensional spherical Gaussians. Querying the target model only four times can capture the majority of the benefits, which enhances the feasibility of the RAPID attack.

\subsection{Model Architecture}

\begin{figure}[t]
    \centering
    \includegraphics[width=0.49\textwidth]{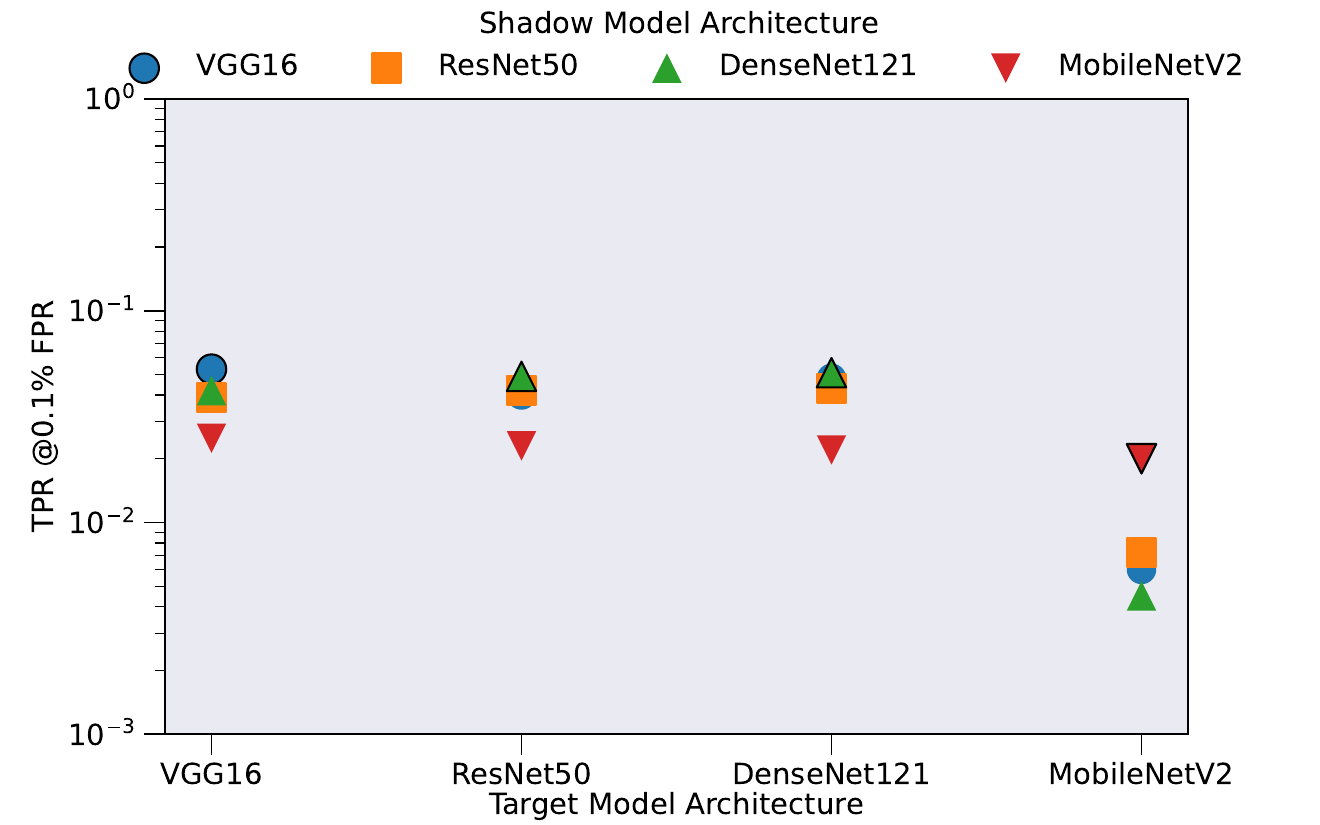}
    \caption{The impact of architecture differences between the target model and the models trained by the adversary (shadow model and reference models) on CINIC-10.}
    \label{Figure architecture}
\end{figure}

Most existing works~\cite{shokri2017membership, yeom2018privacy, sablayrolles2019white, leino2020stolen, ye2022enhanced, liu2022membership, carlini2022membership} have assumed that the adversary has knowledge of the specific architecture of the target model in order to have a larger attack surface. However, this assumption is often not valid. Therefore, we aim to investigate the impact of mismatched model architectures on the attack results. Following the settings of our main experiments, we vary the architectures of the target model, shadow models, and reference models while ensuring consistency in the architectures of shadow models and reference models (which is feasible for the attacker). The experimental results in Figure \ref{Figure architecture} show that the attack performs best when all three models have identical architectures. When the model architectures are completely different, there is only a little drop in attack performance except for MobileNetV2. The phenomenon of degradation is easily understood because the membership score distributions obtained from models with different architectures are significantly different, even if they are trained on the same dataset. This directly leads to our trained \(\mathcal{M}_\text{score}\) incorrectly mapping \(\mathcal{S}(x,y)\) and \(\mathcal{S'}(x,y)\) obtained from \(\mathcal{M}_\text{target}\) and \(\mathcal{M}_\text{refernce}\) to final membership scores. Despite that, our attack still achieves significantly better performance compared to other baseline attacks using the same architecture, as demonstrated in Table \ref{table:vgg16 attack performances}. The notable decrease in attack performance due to the MobileNetV2 architecture can be attributed to the fact that in MobileNetV2, the number of channels in the feature map increases and then decreases, which is contrary to the other three architectures. Remarkably, previous research~\cite{carlini2022membership} has also shown similar experimental results. We hope that future work can provide a clearer explanation for this phenomenon. Overall, our attack demonstrates stronger robustness because it outperforms existing baseline attacks even in more challenging settings, whereas the baseline attacks achieved their results in easier settings.

\subsection{Disjoint Dataset}
\begin{figure}[t]
    \centering
    \subfloat[Balanced Accuracy]{\includegraphics[width=1.7in]{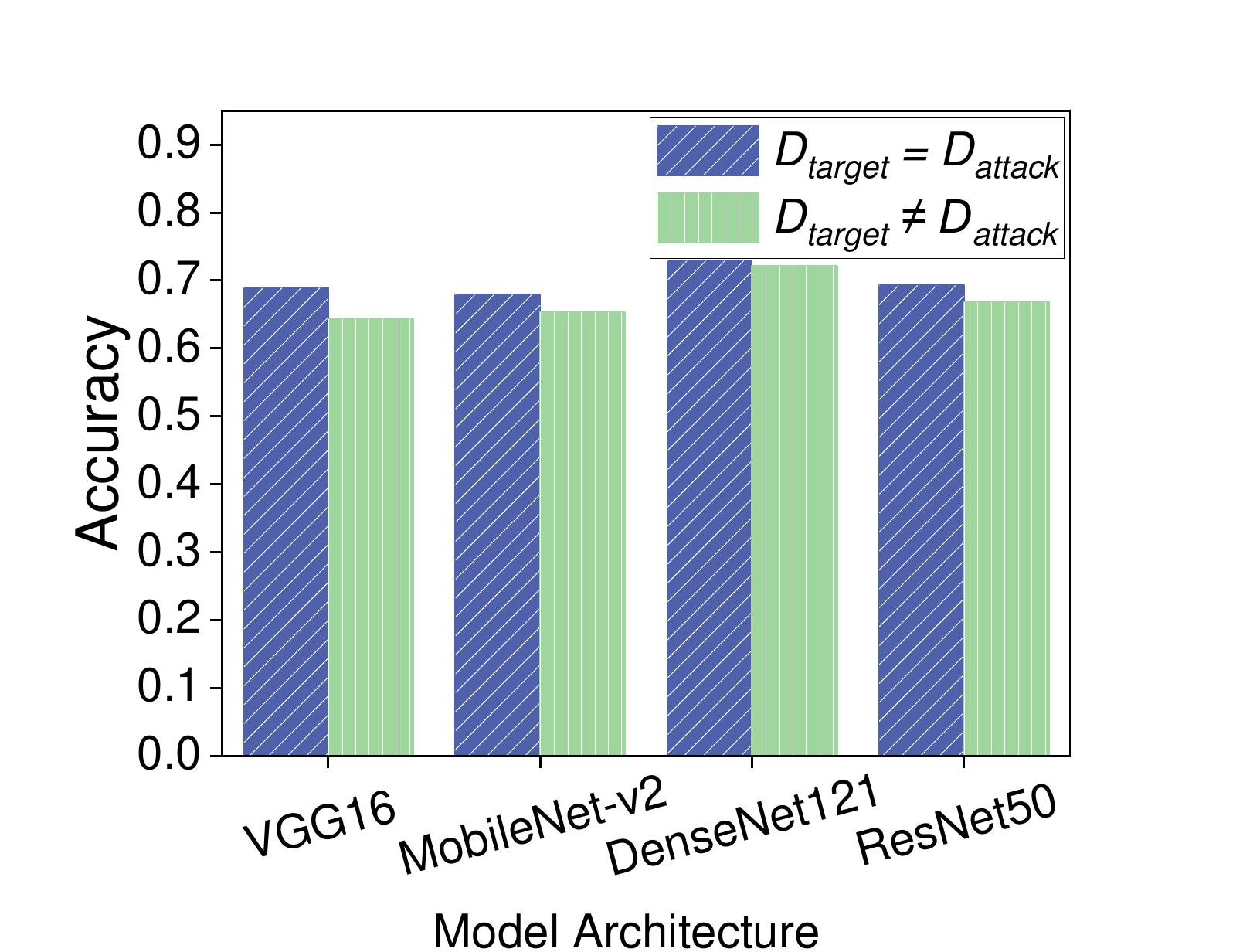} }
    \subfloat[TPR @ 0.1\% FPR]{\includegraphics[width=1.7in]{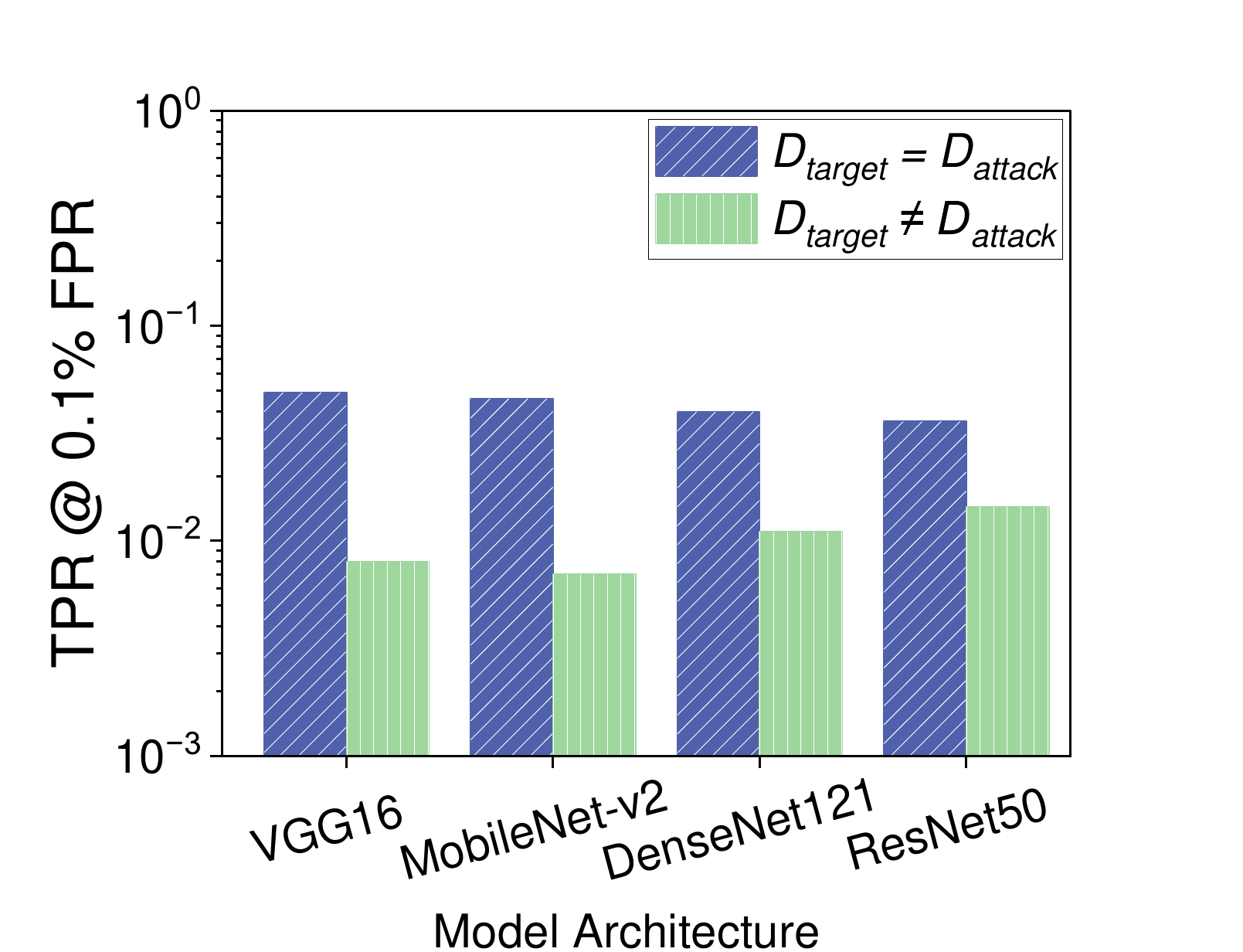} }
    \caption{The impact of distribution shift between the target model training dataset and the attack dataset owned by the adversary (shadow dataset and reference dataset).}
    \label{Figure disjoint}
\end{figure}

In this section, we relax the assumption that the attacker has access to an attack dataset that follows the same distribution as the target model's training dataset. We instead assume that the attacker only has an attack dataset that is disjoint from the target model's training dataset, which they use to train the shadow models and reference models. This is a more realistic condition since it is difficult for the attacker to obtain a dataset that is perfectly aligned with the target training dataset. Specifically, we conduct experiments in the following two settings:
\begin{itemize}\setlength{\itemsep}{0.1em} \setlength{\parskip}{0.1em}
\item \(\mathbbm{D}_{\text{target}}=\mathbbm{D}_{\text{attack}}\). Specifically, we train the target model, shadow model, and reference models using the CIFAR-10 dataset. This setup completely follows the settings in our main experiments.
\item \(\mathbbm{D}_{\text{target}}\neq\mathbbm{D}_{\text{attack}}\). Specifically, we train the target model using a subset of the CIFAR-10 dataset, while we train the shadow model and reference models using the ImageNet portion of the CINIC-10 dataset, following prior work~\cite{liu2022membership, carlini2022membership}.
\end{itemize}
In order to eliminate the influence of overfitting on the attack performance, we keep the same amount of data in both settings. Additionally, the number of queries and reference models remains the same. Figure \ref{Figure disjoint} shows that the distribution shift between \(\mathcal{D}_\text{target}\) and \(\mathcal{D}_\text{attack}\) indeed leads to a noticeable decrease in TPR at 0.1\% FPR.  This is because the decreasing similarity between the shadow model and the target model makes errors in calibrated scores increase, which finally weakens the performance of scoring model on an unseen dataset. Remarkably, our attack still outperforms the majority of baseline attacks in harder settings on Balanced Accuracy.

\begin{table*}[!t]
\centering
\setlength{\tabcolsep}{4.0pt}
\caption{The attack performances of LiRA online version and our RAPID on VGG16 models trained on four benchmark datasets. We use 64 reference models (only OUT models) for RAPID to achieve its optimal performance.}
\scalebox{0.9}
{
\begin{tabular}{l|cccccccccccc}
\toprule
Attack & \multicolumn{4}{c}{TPR @ 0.1\% FPR}&  \multicolumn{4}{c}{AUC}& \multicolumn{4}{c}{Balanced Accuracy}\\
\cmidrule(l{5pt}r{5pt}){2-5}\cmidrule(l{5pt}r{5pt}){6-9}\cmidrule(l{5pt}r{0pt}){10-13}
method& CIFAR-10& CIFAR-100& CINIC-10& SVHN&CIFAR-10& CIFAR-100& CINIC-10& SVHN&CIFAR-10& CIFAR-100& CINIC-10& SVHN\\
\midrule
Carlini et al.~\cite{carlini2022membership}& \textbf{11.9\%} & \textbf{43.6\%} & 12.4\% & \textbf{6.5\%} & 0.790 & 0.972 & 0.778 & 0.629 & 68.9\% & 90.1\% & 63.9\% & 57.4\%\\
Ours&10.9\%&42.3\%&\textbf{13.9\%}&5.8\%&\textbf{0.808}&\textbf{0.974}&\textbf{0.826}&\textbf{0.641}&\textbf{70.5\%}&\textbf{90.6\%}&\textbf{71.8\%}&\textbf{57.9\%}\\
\bottomrule
\end{tabular}
}
\label{tab: compare with lira online}
\end{table*}

\section{Related Work}
\label{sec: 7}
Recently researchers have paid growing emphasis on the importance of high-precision inference in the field of MIAs~\cite{long2020pragmatic, leino2020stolen, sablayrolles2019white, watson2021importance, carlini2022membership, ye2022enhanced, liu2022membership, wen2022canary}. Various attack methods based on difficulty calibration have been proposed to address this challenge. Sablayrolles et al.~\cite{sablayrolles2019white} introduce a method that uses loss from both reference models trained with and without the target point to calibrate the original membership scores. Watson et al.~\cite{watson2021importance} employ a similar approach but replace all reference models with trivial OUT models (trained without the target point). Carlini et al.~\cite{carlini2022membership} take a step further from the aforementioned approaches~\cite{watson2021importance, sablayrolles2019white} by fitting Gaussians to the outputs of the referenced models. It considers the distribution parameters of the target point's loss on a large number of reference models. Ye et al.~\cite{ye2022enhanced} design a model-dependent and sample-dependent attack leveraging distilled models, which are closer to the target model. Liu et al.~\cite{liu2022membership} introduce a Loss Trajectory Attack, which utilizes the distillation trajectory of the target model for membership inference. Wen et al.~\cite{wen2022canary} argue that one limitation of LiRA is that it queries the target model using only the original target data point or its augmentations. They instead learn query vectors that are maximally discriminative; they separate all models trained with the target data point from all models trained without it. In general, previous work has mainly focused on obtaining calibrated scores of higher-quality, while overlooking the impact of attack cost on the practical threat of the attack. Notably, pioneering work by Shokri et al.~\cite{shokri2017membership} also utilizes original outputs, but is primarily based on the intuition that there are differences between outputs from members and non-members. To the best of our knowledge, we are the first to formally utilize the compelling non-member evidence in original outputs to address the inherent errors in difficulty calibration, thereby achieving a more powerful and practical MIA.

\section{Discussion and Limitations}
\label{sec: 8}

\paragraph{Our Paradigm Also Enhances LiRA} 
In addition, our extensive experimental results have demonstrated that RAPID significantly outperforms existing state-of-the-art attacks. However, whether directly re-leveraging original membership scores can enhance more complicated attacks that utilize difficulty calibration (e.g., LiRA) remains an unresolved question. To investigate this question, we do experiments on a VGG16 model trained on the CIFAR-10 dataset. Specifically, we use the concatenation of the original membership scores and the membership scores calculated by LiRA offline (i.e., the results of the one-sided hypothesis test) as features to train a scoring model. The experimental results indicate a significant improvement in LiRA's performance, elevating the TPR @0.1\% FPR from 2.2\% to 4.5\%, the AUC from 0.534 to 0.775, and the Acc from 57.5\% to 68.7\%. In other words, although LiRA has trained numerous reference models to make a Gaussian likelihood estimate, it still potentially misclassifies certain high-loss non-members as members.

\paragraph{RAPID vs. LiRA Online.}
As LiRA online is computationally infeasible, we do not include it as a baseline in our main experiments. However, Carlini et al. \cite{carlini2022membership} provide a clever method to evaluate the theoretical performance of LiRA online. We are interested in whether RAPID could serve as a practical alternative to LiRA online in real-world scenarios. Specifically, they combine members and non-members into a set, and then randomly sample half of the data to train a reference model. This process is repeated 256 times. For any given target sample, since it has a 50\% chance of being sampled into the training set of any reference model, there are approximately 128 reference models serving as its IN models, and 128 reference models serving as its OUT models. Although this method circumvents the necessity to train 128 IN models for each target sample, it remains impractical in reality because the adversary cannot access all potential member samples before the inference time. Furthermore, this implementation potentially boosts LiRA online's attack performance compared to traditional methods (i.e., those used in our main experiments)---each IN/OUT model shares about half of its training data with the target model, making the IN/OUT models highly similar to the target model. This similarity makes calibrated membership scores more accurate, as they rely more on membership status rather than model parameters (model characteristics). To ensure a fair comparison, we train the reference models for RAPID using a similar method to LiRA. However, note that RAPID remains a practical offline attack, with the key difference being that the reference models and the target model share some training data. As demonstrated in Table \ref{tab: compare with lira online}, our RAPID, as an offline attack, achieves nearly the same TPR at 0.1\% FPR as LiRA online, along with higher AUC and balanced accuracy results. Consequently, while LiRA online is not so practical in real-world scenarios, our RAPID represents an equivalent alternative that is computationally feasible. 

\paragraph{Limitations.} 
Our work has several limitations. First, the effectiveness of all existing MIAs mainly relies on identifying out-of-distribution member samples (i.e., samples only receiving high membership scores from the target model). This, to some extent, limits the performance of MIAs. Although our RAPID achieves state-of-the-art performance, it does not fully address this issue. Second, we only evaluate RAPID on public datasets, and its effectiveness and sensitivity to specific populations (or subgroups of datasets) have not been fully investigated. Third, our evaluation on LLMs is limited to masked language models, and the performance of RAPID on autoregressive language models has not been studied. We will conduct experiments on more LLMs in the future. Despite these limitations, we believe our study provides insight into the limitations of difficulty calibration, issues of MIA practicality, and finally the potential solutions to the aforementioned issues. 

\section{Conclusion}

In this paper, we have emphasized that existing reference-based MIAs do not fully utilize the non-member evidence contained in the original membership scores, which can be re-leveraged to correct the misclassification of non-members caused by difficulty calibration. Therefore, we have introduced a new attack RAPID, which directly corrects the inherent errors in difficulty calibration by training a scoring model to map the original membership scores and the calibrated scores to the final membership scores. This improves the attack efficacy by eliminating the need for: 1) training a large number of models; and 2) near-unlimited query access to the target model. Extensive experiments demonstrate the state-of-the-art performance of RAPID in both classic image domains and recent fields of LLMs. We hope our research can advance the development of more efficacious techniques for quantifying privacy loss and protecting data privacy.

\begin{acks}
We thank the anonymous reviewers for their constructive comments. This work is supported in part by the National Natural Science Foundation of China under Grants No. 61872430, 61402342, and 61772384. Additionally, it is sponsored by funding from Research on Attacks and Defenses in Split Learning by Ant Group, China. Any opinions, findings, and conclusions expressed in this paper are those of the authors only and do not necessarily reflect the views of any funding agencies.
\end{acks}

\bibliographystyle{ACM-Reference-Format}
\balance
\bibliography{sample-base}

\clearpage
\onecolumn
\appendix

\section{Data Splits on Different Datasets}
\label{sec:data split}

\begin{table*}[h]
\newcommand{\tabincell}[2]{\begin{tabular}{@{}#1@{}}#2\end{tabular}}
\centering
\setlength{\tabcolsep}{2.5pt}
\caption{Data splits of all the datasets used in our main experiments.}
\scalebox{1.0}
{
\begin{tabular}{l|cccccc}
\toprule
Dataset& $\mathcal{D}^t_{train}$& $\mathcal{D}^t_{test}$ & $\mathcal{D}^s_{train}$& $\mathcal{D}^s_{test}$&$\mathcal{D}^r_{train}$& $\mathcal{D}^r_{test}$\\
\midrule
CIFAR-10& 10000& 10000& 10000& 10000& 10000& 10000\\
CIFAR-100& 10000& 10000& 10000& 10000& 10000& 10000\\
CINIC-10& 45000& 45000& 45000& 45000& 45000& 45000\\
SVHN& 16548& 16548& 16548& 16548& 16548&16548\\
Location& 835& 835& 835& 835& 835& 835\\
Texas& 11222& 11222& 11222& 11222& 11222& 11222\\
cola& 1513& 1513& 1513& 1513& 1513& 1513\\
cb& 51& 51& 51& 51& 51& 51\\
mrpc& 968& 968& 968& 968& 968& 968\\
\bottomrule
\end{tabular}
}
\label{table:Multiple}
\end{table*}

\section{Additional Experimental Results on Other Models}
\label{Additional Experimental Results on Other Models}

It is worth noting that due to the high computational cost of LiRA and Canary, when evaluating them on other model architectures, we randomly selected a benchmark dataset for each architecture. When evaluating other advanced attacks, we utilized all four benchmark datasets for each architecture.

\begin{table*}[!h]
\centering
\setlength{\tabcolsep}{4.0pt}
\caption{The attack performances of different attacks on ResNet50 models trained on four benchmark datasets.}
\scalebox{0.75}
{
\begin{tabular}{l|cccccccccccc}
\toprule
Attack & \multicolumn{4}{c}{TPR @ 0.1\% FPR}&  \multicolumn{4}{c}{AUC}& \multicolumn{4}{c}{Balanced Accuracy}\\
\cmidrule(l{5pt}r{5pt}){2-5}\cmidrule(l{5pt}r{5pt}){6-9}\cmidrule(l{5pt}r{0pt}){10-13}
method& CIFAR-10& CIFAR-100& CINIC-10& SVHN&CIFAR-10& CIFAR-100& CINIC-10& SVHN&CIFAR-10& CIFAR-100& CINIC-10& SVHN\\
\midrule
Yeom et al. \cite{yeom2018privacy}&0.0\%&0.4\%&0.1\%&0.0\%&0.655&0.934&0.669&0.539&64.2\%&90.4\%&65.4\%&54.6\%\\
Yuan et al. \cite{yuan2022membership}&0.2\%&3.2\%&0.2\%&0.2\%&0.687&0.950&0.688&0.550&64.6\%&90.5\%&65.7\%&54.7\%\\
Watson et al. \cite{watson2021importance}&1.2\%&2.4\%&1.4\%&0.9\%&0.650&0.783&0.644&0.560&60.7\%&71.0\%&59.3\%&53.3\%\\
Ye et al. \cite{ye2022enhanced}&0.0\%&0.0\%&0.8\%&0.6\%&0.641&0.747&0.639&0.561&52.0\%&70.9\%&59.3\%&53.2\%\\
Liu et al. \cite{liu2022membership}&0.0\%&6.5\%&1.3\%&0.7\%&0.745&0.970&0.745&0.589&67.2\%&91.7\%&66.4\%&55.1\%\\
\midrule
Ours&\textbf{3.3\%}&\textbf{27.4\%}&\textbf{3.9\%}&\textbf{2.7\%}&\textbf{0.779}&\textbf{0.984}&\textbf{0.792}&\textbf{0.607}&\textbf{69.6\%}&\textbf{94.0\%}&\textbf{70.0\%}&\textbf{56.5\%}\\
\bottomrule
\end{tabular}
}
\label{table:resnet attack performances}
\end{table*}

\begin{table*}[!h]
\centering
\setlength{\tabcolsep}{4.0pt}
\caption{The attack performances of different attacks on MobileNetV2 models trained on four benchmark datasets.}
\scalebox{0.75}
{
\begin{tabular}{l|cccccccccccc}
\toprule
Attack & \multicolumn{4}{c}{TPR @ 0.1\% FPR}&  \multicolumn{4}{c}{AUC}& \multicolumn{4}{c}{Balanced Accuracy}\\
\cmidrule(l{5pt}r{5pt}){2-5}\cmidrule(l{5pt}r{5pt}){6-9}\cmidrule(l{5pt}r{0pt}){10-13}
method& CIFAR-10& CIFAR-100& CINIC-10& SVHN&CIFAR-10& CIFAR-100& CINIC-10& SVHN&CIFAR-10& CIFAR-100& CINIC-10& SVHN\\
\midrule
Yeom et al. \cite{yeom2018privacy}&0.0\%&0.0\%&0.0\%&0.2\%&0.625&0.866&0.574&0.560&62.8\%&84.6\%&57.8\%&55.5\%\\
Yuan et al. \cite{yuan2022membership}&0.1\%&1.7\%&0.2\%&0.1\%&0.659&0.901&0.604&0.556&63.0\%&84.9\%&58.2\%&55.3\%\\
Watson et al. \cite{watson2021importance}&0.9\%&1.9\%&0.9\%&1.0\%&0.634&0.738&0.624&0.570&59.5\%&69.0\%&58.2\%&53.8\%\\
Ye et al. \cite{ye2022enhanced}&0.6\%&1.0\%&0.5\%&0.6\%&0.617&0.701&0.623&0.572&58.0\%&69.3\%&58.3\%&54.0\%\\
Liu et al. \cite{liu2022membership}&1.1\%&6.6\%&0.9\%&0.9\%&0.710&0.940&0.657&0.600&64.3\%&86.4\%&60.1\%&56.3\%\\
\midrule
Ours&\textbf{4.1\%}&\textbf{16.9\%}&\textbf{2.0\%}&\textbf{2.5\%}&\textbf{0.760}&\textbf{0.959}&\textbf{0.686}&\textbf{0.619}&\textbf{67.6\%}&\textbf{89.3\%}&\textbf{62.0\%}&\textbf{57.2\%}\\
\bottomrule
\end{tabular}
}
\label{table:mobilenetv2 attack performances}
\end{table*}

\begin{table*}[!h]
\centering
\setlength{\tabcolsep}{4.0pt}
\caption{The attack performances of different attacks on DenseNet121 models trained on four benchmark datasets.}
\scalebox{0.75}
{
\begin{tabular}{l|cccccccccccc}
\toprule
Attack & \multicolumn{4}{c}{TPR @ 0.1\% FPR}&  \multicolumn{4}{c}{AUC}& \multicolumn{4}{c}{Balanced Accuracy}\\
\cmidrule(l{5pt}r{5pt}){2-5}\cmidrule(l{5pt}r{5pt}){6-9}\cmidrule(l{5pt}r{0pt}){10-13}
method& CIFAR-10& CIFAR-100& CINIC-10& SVHN&CIFAR-10& CIFAR-100& CINIC-10& SVHN&CIFAR-10& CIFAR-100& CINIC-10& SVHN\\
\midrule
Yeom et al. \cite{yeom2018privacy}&0.1\%&0.0\%&0.0\%&0.1\%&0.688&0.929&0.670&0.571&68.3\%&91.1\%&67.3\%&56.2\%\\
Yuan et al. \cite{yuan2022membership}&0.1\%&3.6\%&0.1\%&0.1\%&0.729&0.955&0.709&0.558&68.9\%&91.5\%&67.5\%&55.6\%\\
Watson et al. \cite{watson2021importance}&1.0\%&2.2\%&1.0\%&1.2\%&0.633&0.785&0.650&0.571&59.8\%&71.5\%&61.1\%&53.5\%\\
Ye et al. \cite{ye2022enhanced}&0.9\%&0.9\%&0.8\%&0.9\%&0.636&0.768&0.648&0.571&60.7\%&71.2\%&60.1\%&53.8\%\\
Liu et al. \cite{liu2022membership}&0.8\%&11.0\%&2.3\%&1.2\%&0.776&0.972&0.775&0.603&69.6\%&91.8\%&68.7\%&56.5\%\\
\midrule
Ours&\textbf{3.4\%}&\textbf{28.2\%}&\textbf{5.4\%}&\textbf{2.7\%}&\textbf{0.805}&\textbf{0.983}&\textbf{0.810}&\textbf{0.624}&\textbf{72.3\%}&\textbf{93.4\%}&\textbf{69.1\%}&\textbf{58.0\%}\\
\bottomrule
\end{tabular}
}
\label{table:densenet121 attack performances}
\end{table*}
\clearpage
\begin{figure*}[!h]
    \centering
    \subfloat[CIFAR-10]{\includegraphics[width=1.7in]{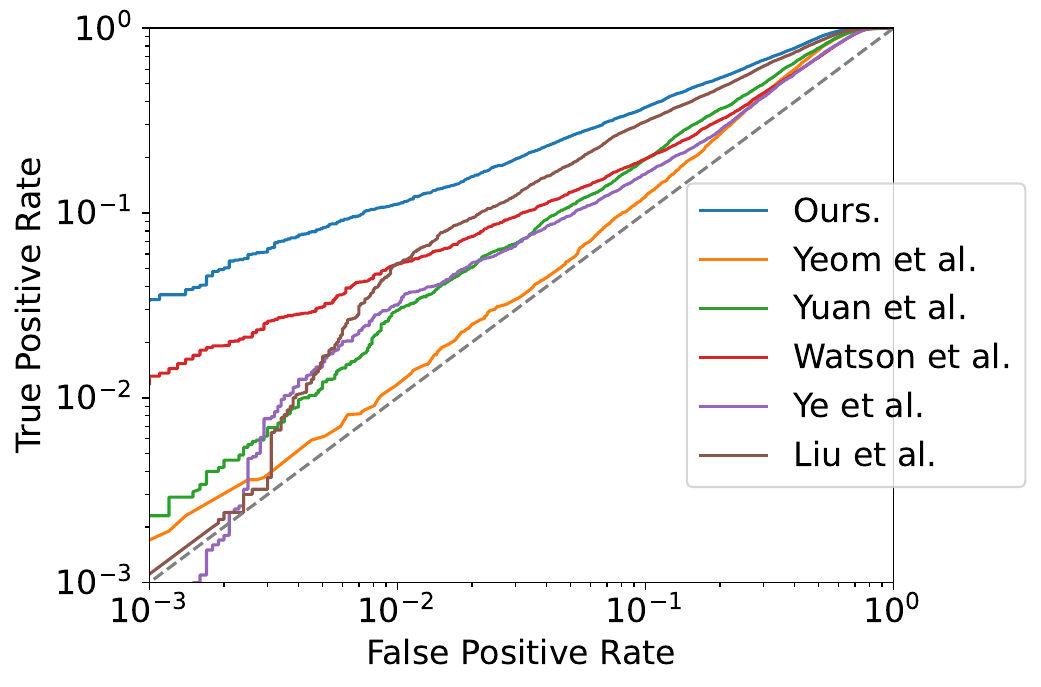} }
    \subfloat[CIFAR-100]{\includegraphics[width=1.7in]{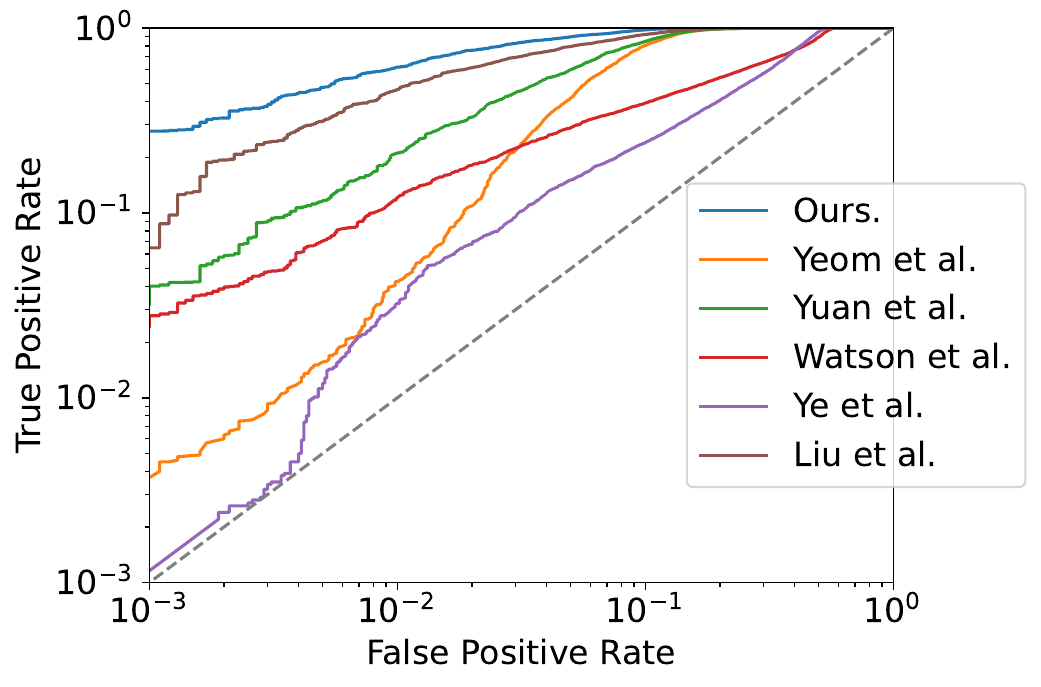} }
    \subfloat[CINIC-10]{\includegraphics[width=1.7in]{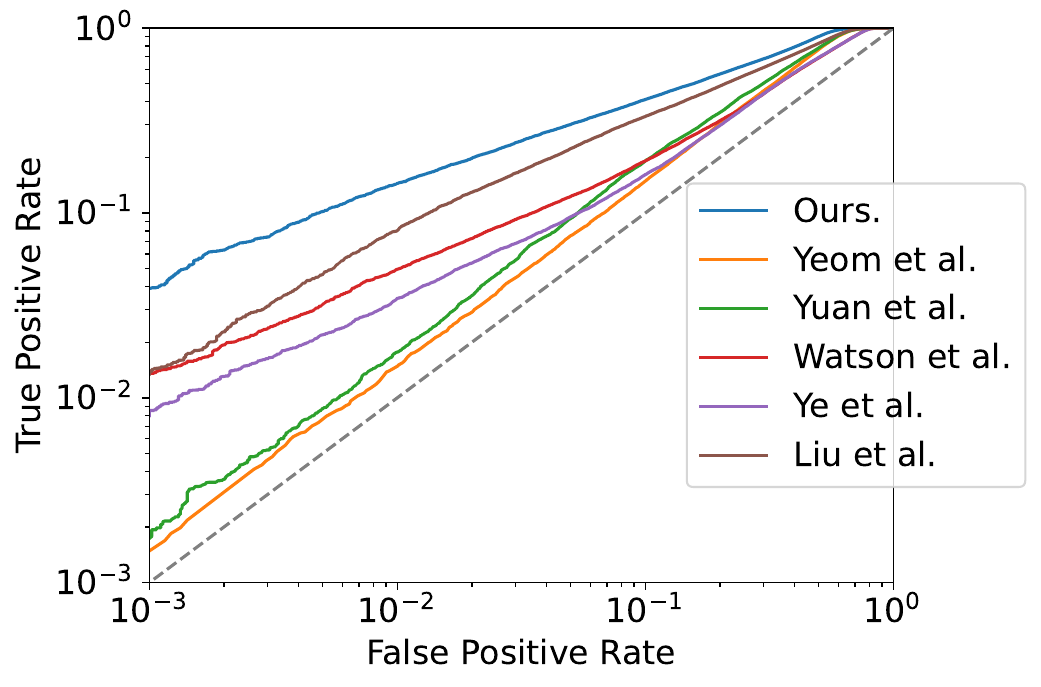} }
    \subfloat[SVHN]{\includegraphics[width=1.7in]{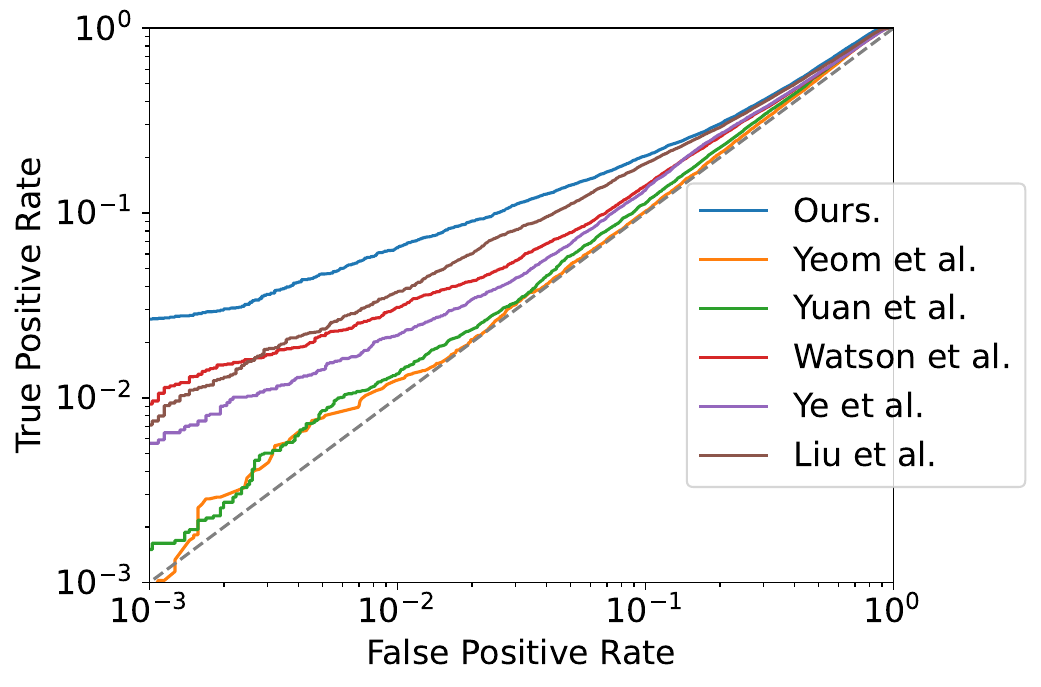} }
    \caption{The ROC curves of attack results on ResNet50 models trained on four benchmark datasets.}
\label{Figure 12}
\end{figure*}
\begin{figure*}[!h]
    \centering
    \subfloat[CIFAR-10]{\includegraphics[width=1.7in]{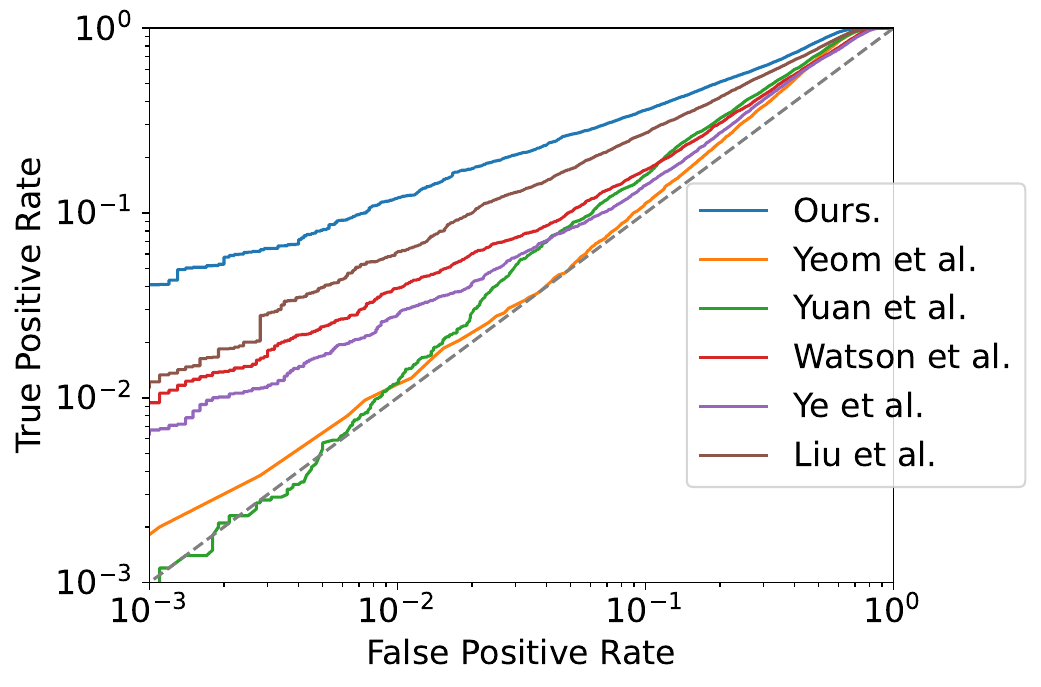} }
    \subfloat[CIFAR-100]{\includegraphics[width=1.7in]{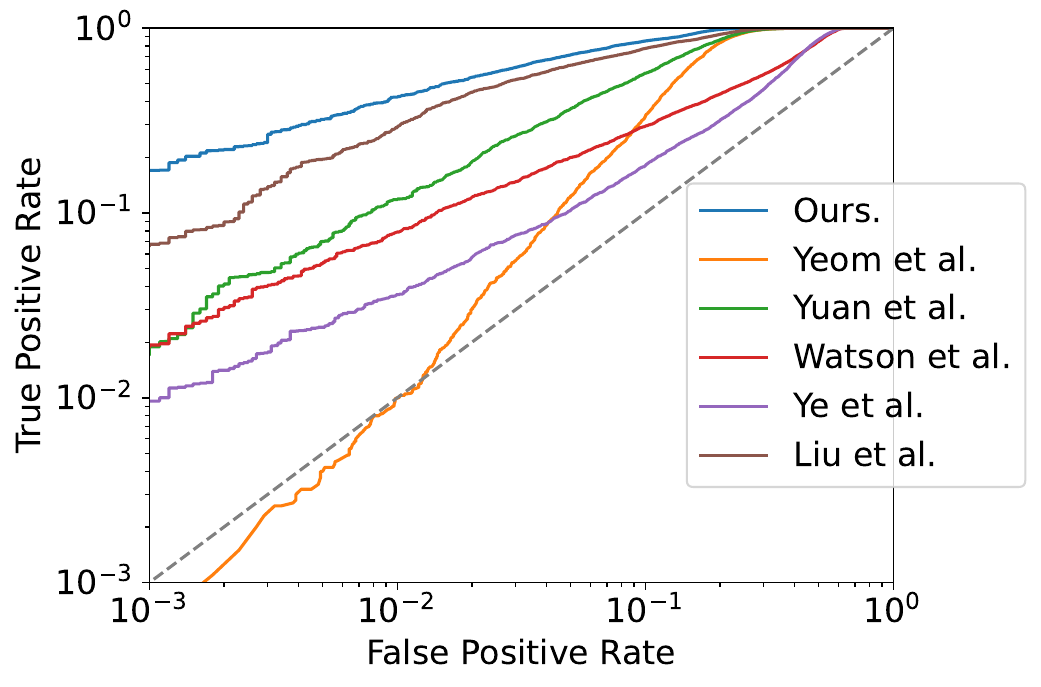} }
    \subfloat[CINIC-10]{\includegraphics[width=1.7in]{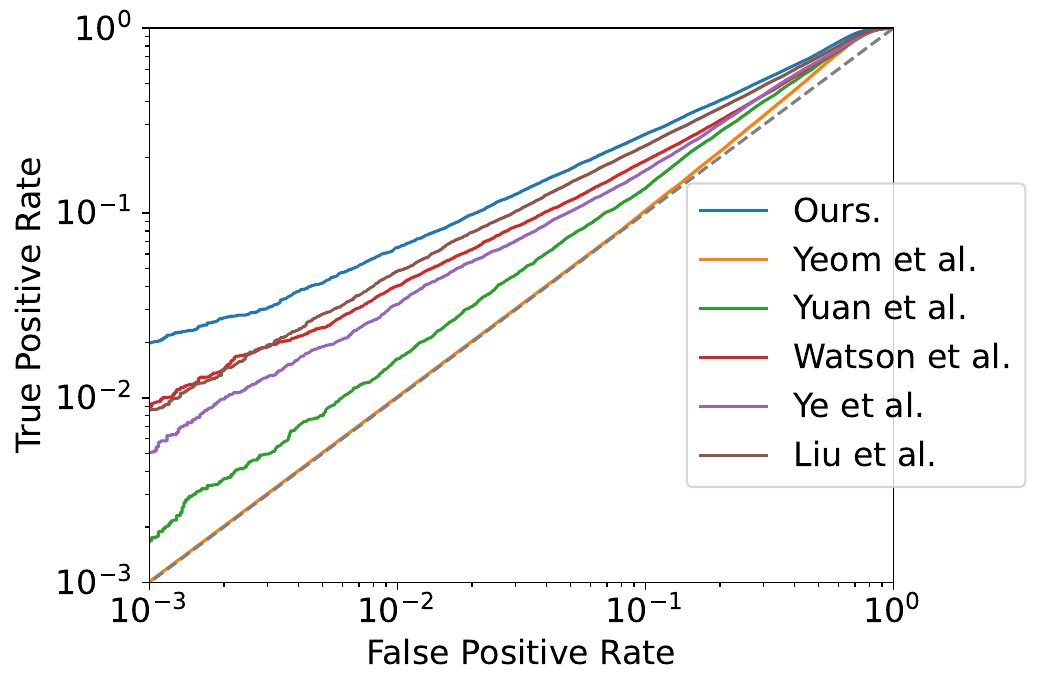} }
    \subfloat[SVHN]{\includegraphics[width=1.7in]{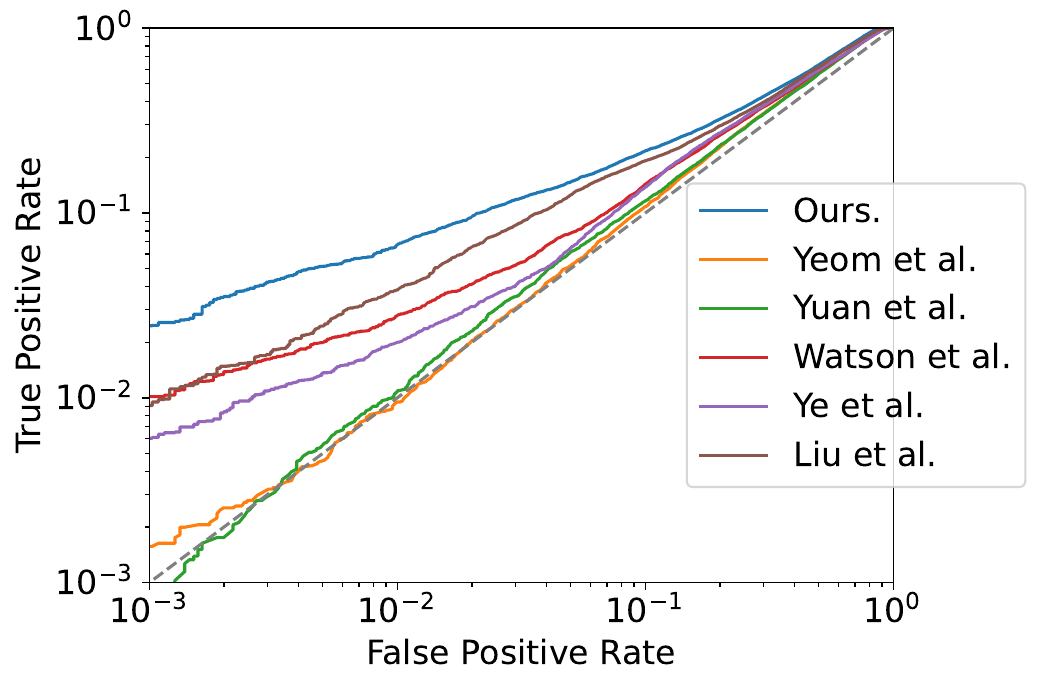} }
    \caption{The ROC curves of attack results on MobileNetV2 models trained on four benchmark datasets.}
\label{Figure 13}
\end{figure*}
\begin{figure*}[!h]
    \centering
    \subfloat[CIFAR-10]{\includegraphics[width=1.7in]{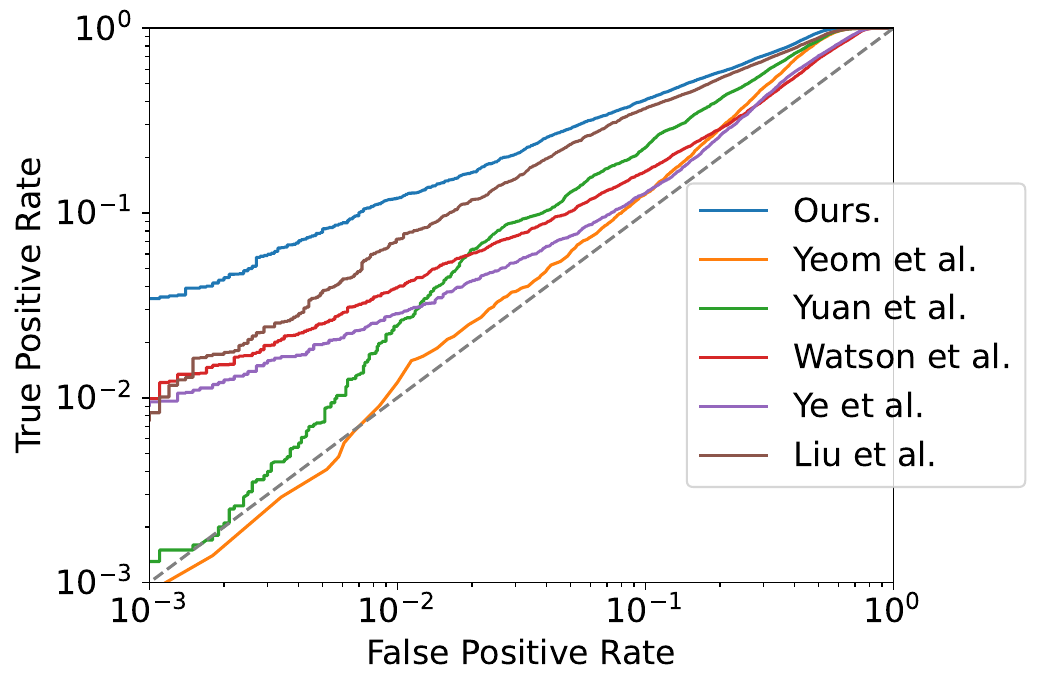} }
    \subfloat[CIFAR-100]{\includegraphics[width=1.7in]{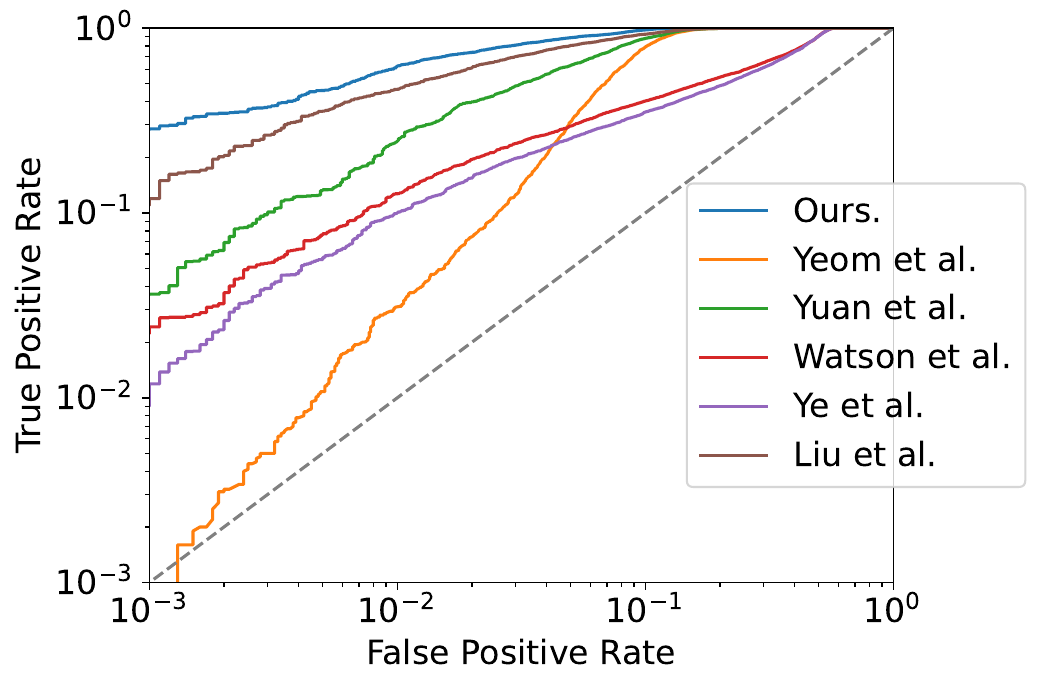} }
    \subfloat[CINIC-10]{\includegraphics[width=1.7in]{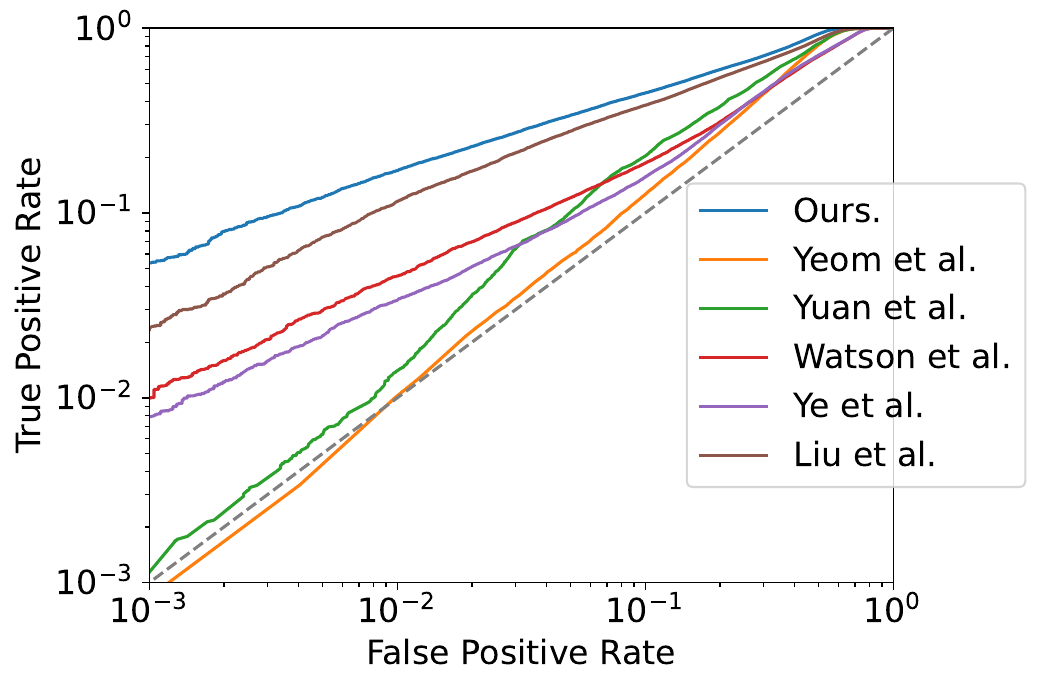} }
    \subfloat[SVHN]{\includegraphics[width=1.7in]{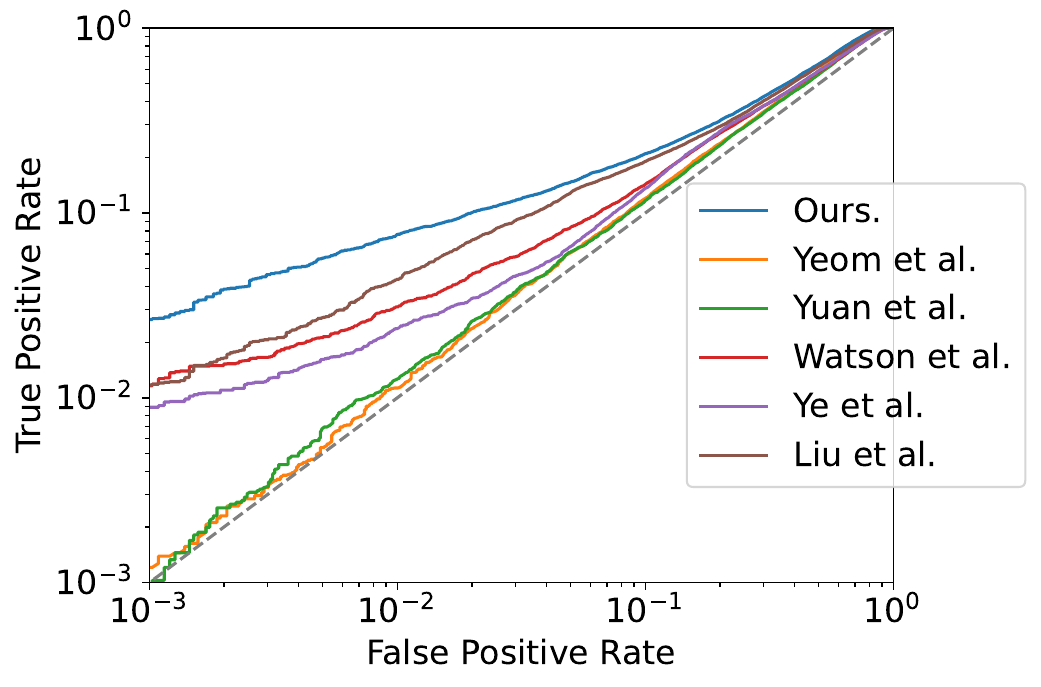} }
    \caption{The ROC curves of attack results on DenseNet121 models trained on four benchmark datasets.}
\label{Figure 14}
\end{figure*}

\clearpage

\begin{table*}[!t]
\centering
\setlength{\tabcolsep}{4.0pt}
\caption{The attack performances of different attacks including LiRA and Canary on different model architectures and benchmark datasets.}
\scalebox{0.8}
{
\begin{tabular}{l|ccccccccc}
\toprule
 & \multicolumn{3}{c}{TPR @ 0.1\% FPR}&  \multicolumn{3}{c}{AUC}& \multicolumn{3}{c}{Balanced Accuracy}\\
\cmidrule(l{5pt}r{5pt}){2-4}\cmidrule(l{5pt}r{5pt}){5-7}\cmidrule(l{5pt}r{0pt}){8-10}
Attack& CIFAR-10& CIFAR-100& SVHN&CIFAR-10& CIFAR-100& SVHN&CIFAR-10& CIFAR-100& SVHN\\
method& MobileNetV2& DenseNet121& ResNet50&MobileNetV2& DenseNet121& ResNet50&MobileNetV2& DenseNet121& ResNet50\\
\midrule
Yeom et al.~\cite{yeom2018privacy}&0.0\%&0.0\%&0.0\%&0.625&0.929&0.539&62.8\%&91.1\%&54.6\%\\
Yuan et al.~\cite{yuan2022membership}&0.1\%&3.6\%&0.2\%&0.659&0.955&0.550&63.0\%&91.5\%&54.7\%\\
Watson et al.~\cite{watson2021importance}&1.0\%&2.2\%&0.9\%&0.634&0.785&0.560&59.5\%&71.5\%&53.3\%\\
Carlini et al.~\cite{carlini2022membership}&2.7\%&25.1\%&1.3\%&0.543&0.900&0.499&57.3\%&84.4\%&51.7\%\\
Ye et al.~\cite{ye2022enhanced}&0.6\%&1.0\%&0.6\%&0.617&0.768&0.561&58.0\%&71.2\%&53.2\%\\
Liu et al.~\cite{liu2022membership}&1.1\%&11.0\%&0.7\%&0.710&0.972&0.589&64.3\%&91.8\%&55.1\%\\
Wen et al.~\cite{wen2022canary}&0.1\%&12.7\%&0.9\%&0.499&0.899&0.522&50.8\%&83.6\%&52.7\%\\
\midrule
Ours&\textbf{4.1\%}&\textbf{28.2\%}&\textbf{2.7\%}&\textbf{0.760}&\textbf{0.983}&\textbf{0.607}&\textbf{67.6\%}&\textbf{93.4\%}&\textbf{56.5\%}\\
\bottomrule
\end{tabular}
}
\label{table:attack performances lira}
\end{table*}

\begin{figure*}[ht]
    \centering
    \subfloat[CIFAR-10 \& MobileNetV2]{\includegraphics[width=1.7in]{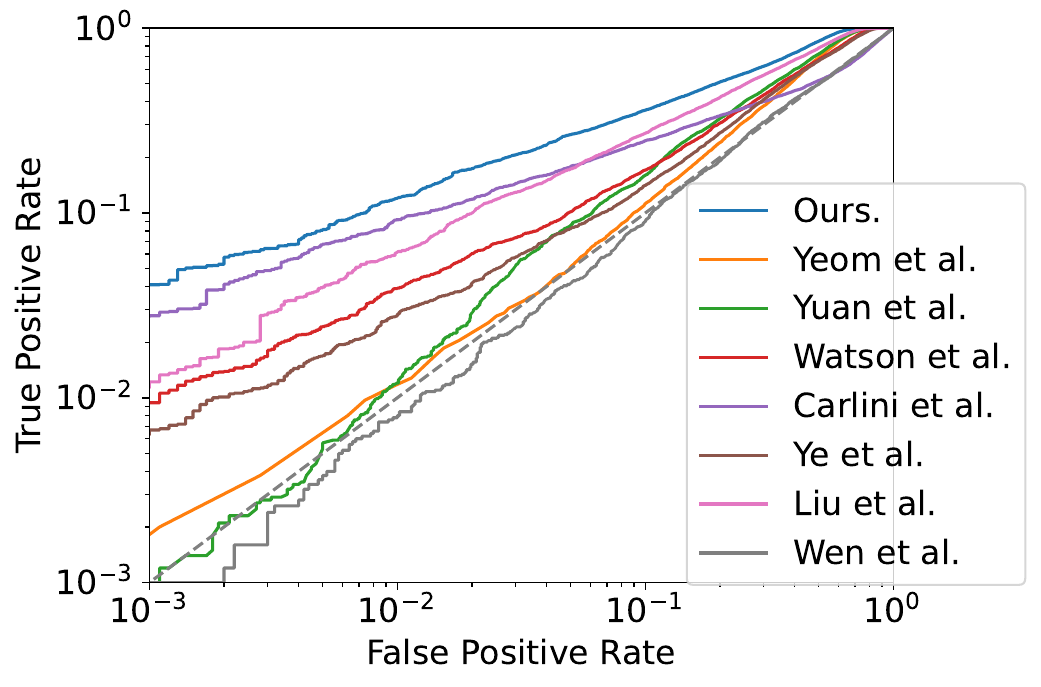} }
    \subfloat[CIFAR-100 \& DenseNet121]{\includegraphics[width=1.7in]{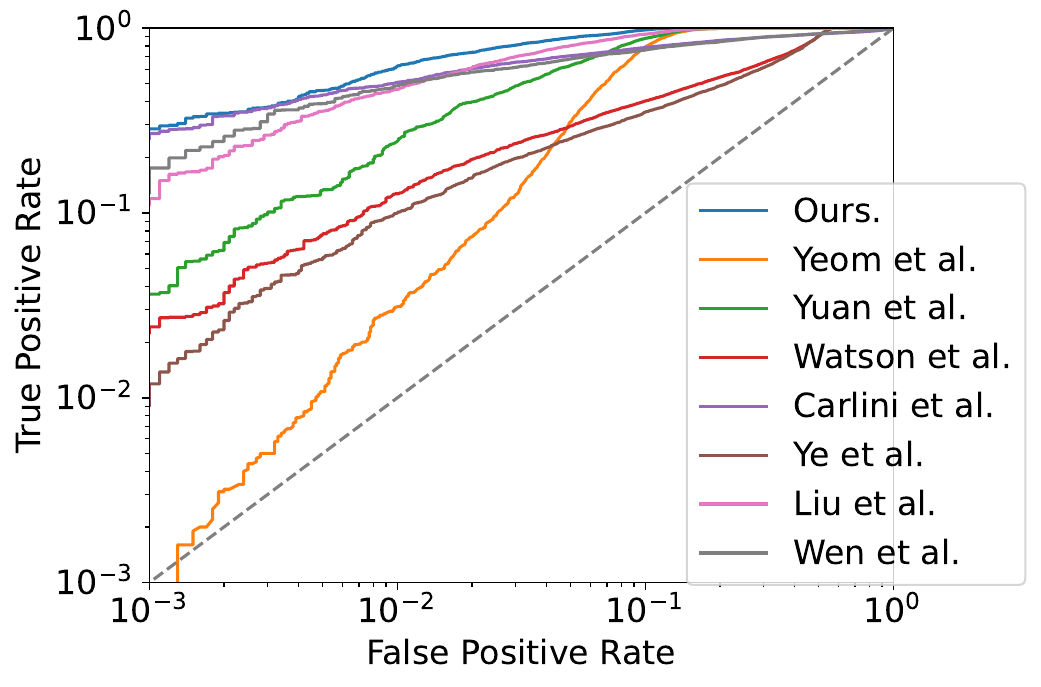} }
    \subfloat[SVHN \& ResNet50]{\includegraphics[width=1.7in]{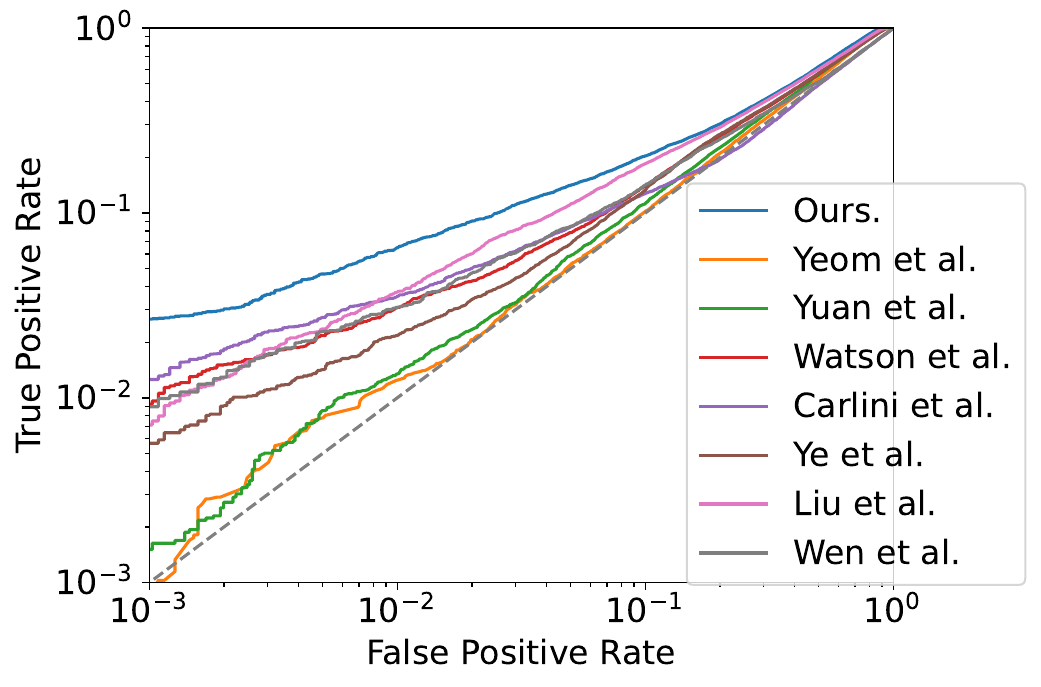} }
    \caption{The ROC curves of different attacks including LiRA and Canary on different model architectures and benchmark datasets.}
\label{Figure: ROC Curves with lira}
\end{figure*}

\section{Additional Experimental Results on Location and Texas Datasets}
\label{Additional Experimental Results on Location and Texas Datasets}

Experimental results on Location and Texas datasets do not include an evaluation of Canary because Canary cannot be directly applied to attack discrete data, such as text or tabular data~\cite{wen2022canary}.
\begin{table*}[!h]
\centering
\setlength{\tabcolsep}{4.0pt}
\caption{The attack performances of different attacks on a 2-layer MLP trained on Location and Texas.}
\scalebox{1.0}
{
\begin{tabular}{l|cccccc}
\toprule
Attack & \multicolumn{2}{c}{TPR @ 0.1\% FPR}&  \multicolumn{2}{c}{AUC}& \multicolumn{2}{c}{Balanced Accuracy}\\
\cmidrule(l{5pt}r{5pt}){2-3}\cmidrule(l{5pt}r{5pt}){4-5}\cmidrule(l{5pt}r{0pt}){6-7}
method& Location& Texas& Location& Texas& Location& Texas\\
\midrule
Yeom et al. \cite{yeom2018privacy}&0.2\%&0.1\%&0.870&0.776&82.2\%&72.5\%\\
Yuan et al. \cite{yuan2022membership}&0.7\%&1.5\%&0.895&0.818&81.8\%&74.1\%\\
Watson et al. \cite{watson2021importance}&0.4\%&4.7\%&0.817&0.760&74.9\%&67.9\%\\
Carlini et al. \cite{carlini2022membership}&9.0\%&4.3\%&0.839&0.681&76.9\%&65.0\%\\
Ye et al. \cite{ye2022enhanced}&3.0\%&1.5\%&0.806&0.691&72.2\%&62.7\%\\
Liu et al. \cite{liu2022membership}&2.6\%&2.5\%&\textbf{0.938}&0.818&\textbf{87.1}\%&74.7\%\\
\midrule
Ours&\textbf{16.4\%}&\textbf{7.7\%}&0.932&\textbf{0.864}&85.3\%&\textbf{77.1\%}\\
\bottomrule
\end{tabular}
}
\label{table:location&Texas attack performances}
\end{table*}
\clearpage
\begin{figure*}[!h]
    \centering
    \subfloat[Location]{\includegraphics[width=2.7in]{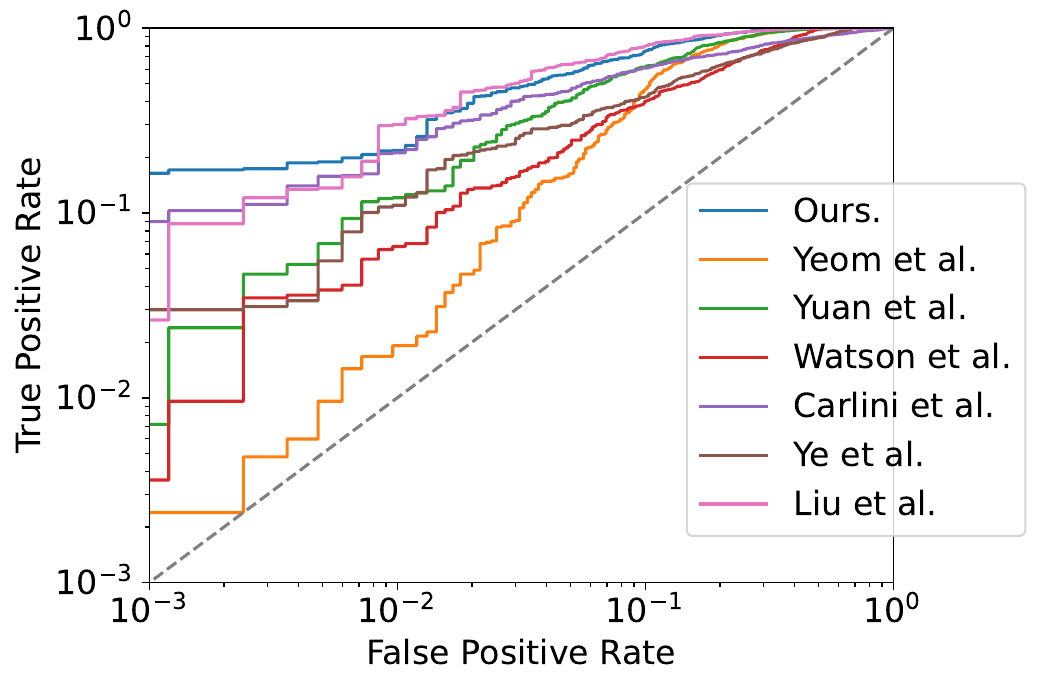} }
    \subfloat[Texas]{\includegraphics[width=2.7in]{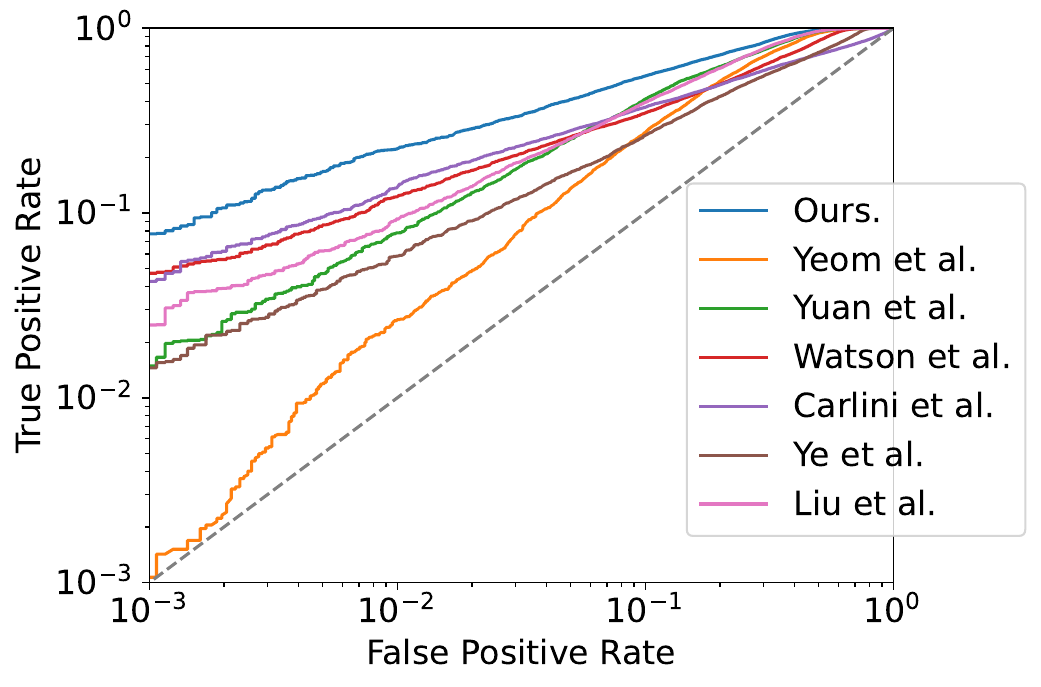} }
    \caption{The ROC curves of attack results on a 2-layer MLP trained on Location and Texas.}
\label{Figure 15}
\end{figure*}

\section{Attacking DP-SGD Using Different $\sigma$}
\label{sec:Attacking DP-SGD}
\begin{figure*}[!h]
    \centering
    \subfloat[\(\sigma = 0.0\)]{\includegraphics[width=3.4in]{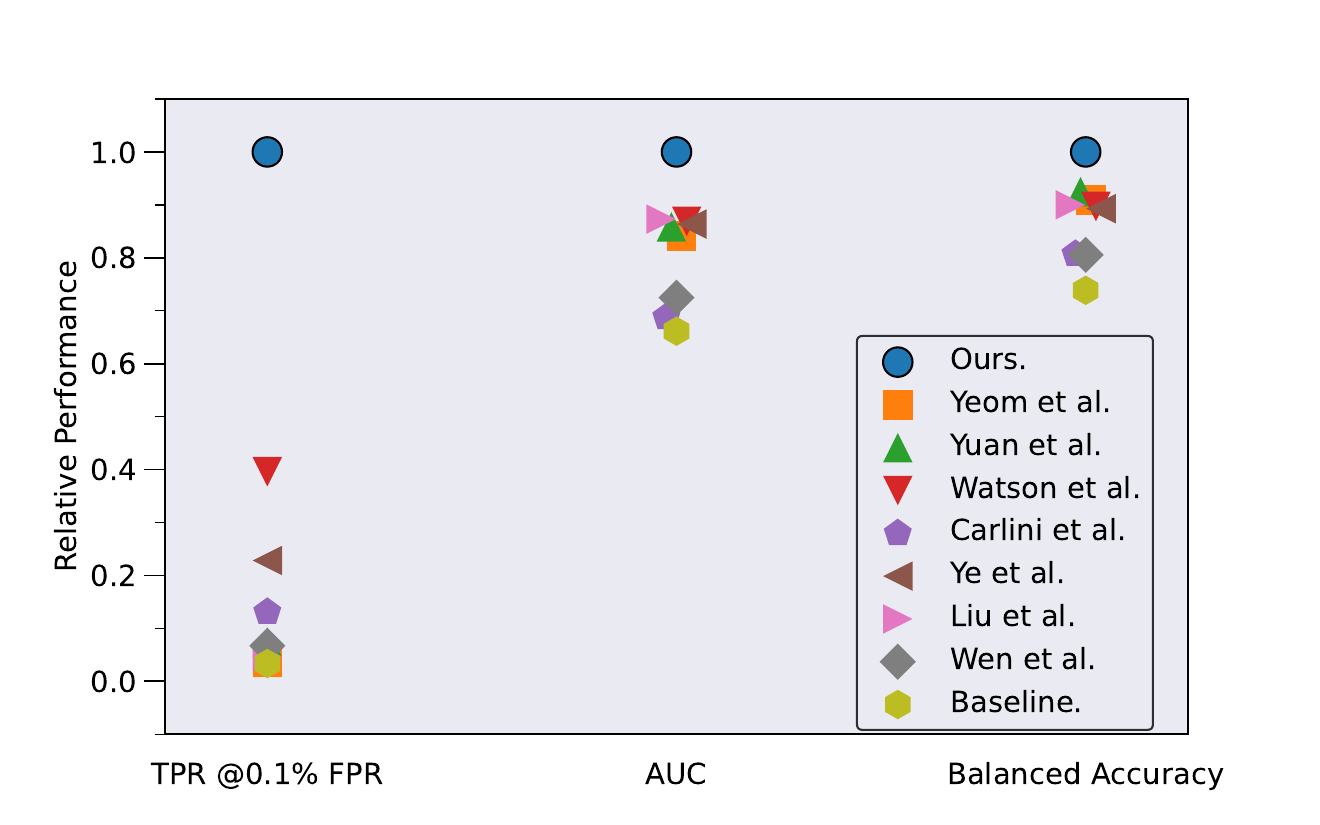} }
    \subfloat[\(\sigma = 0.2\)]{\includegraphics[width=3.4in]{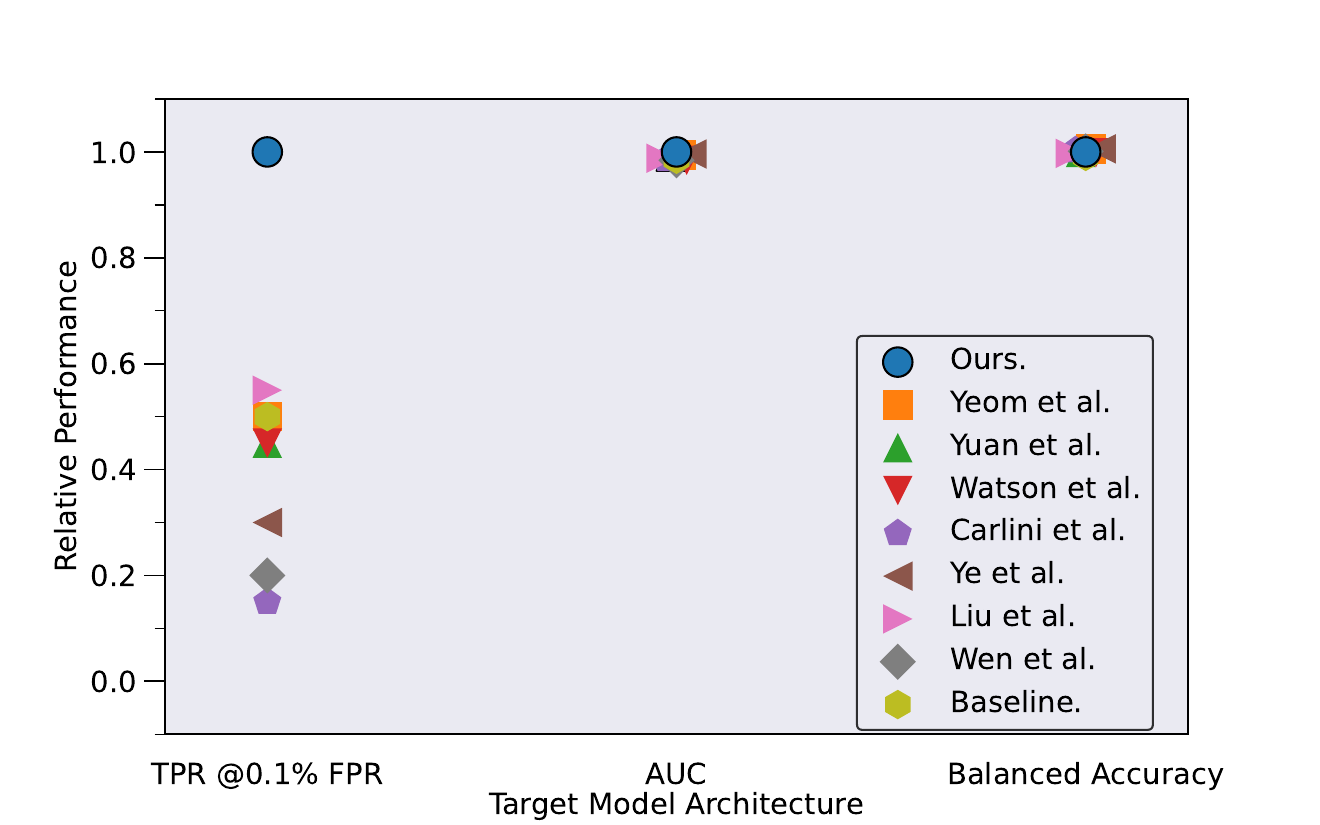} }  \\
    \subfloat[\(\sigma = 0.5\)]{\includegraphics[width=3.4in]{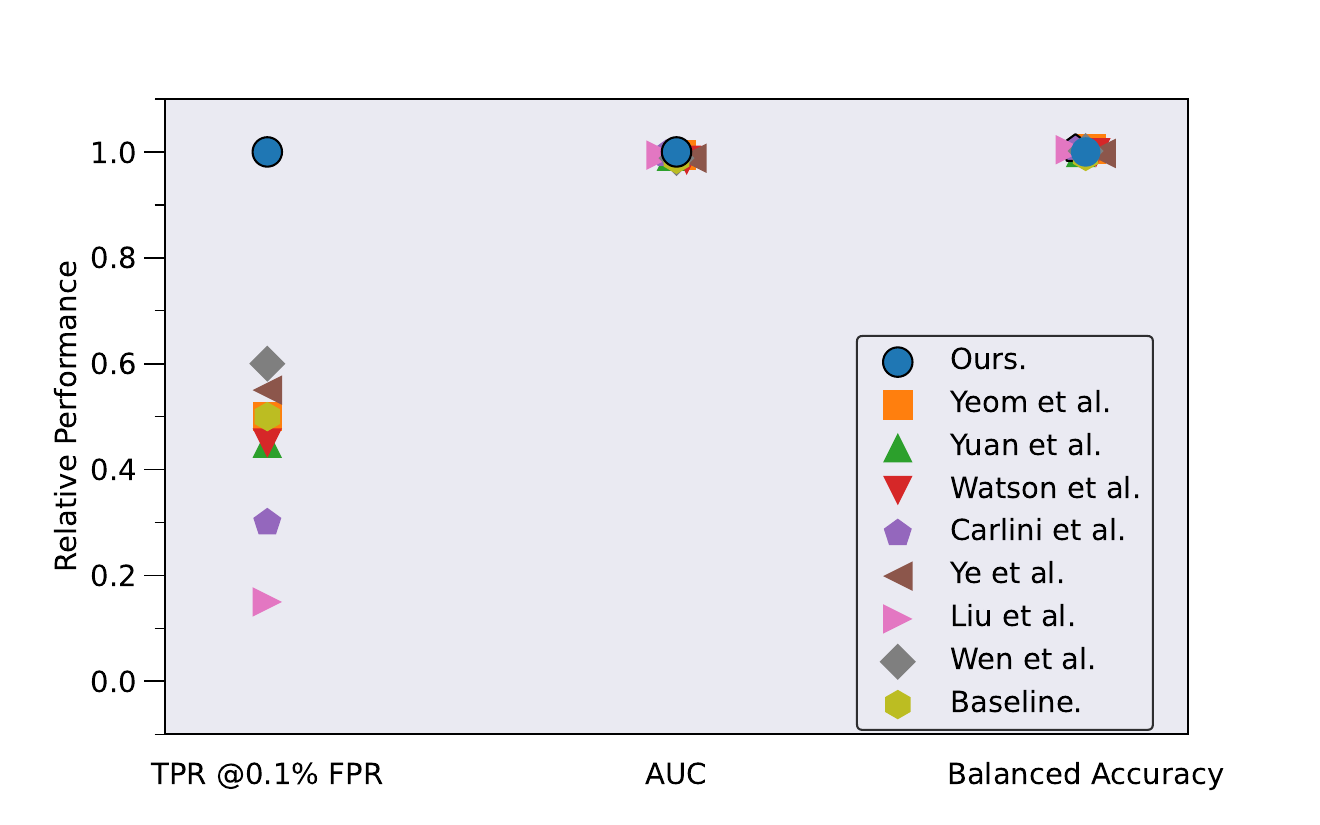} }
    \subfloat[\(\sigma = 1.0\)]{\includegraphics[width=3.4in]{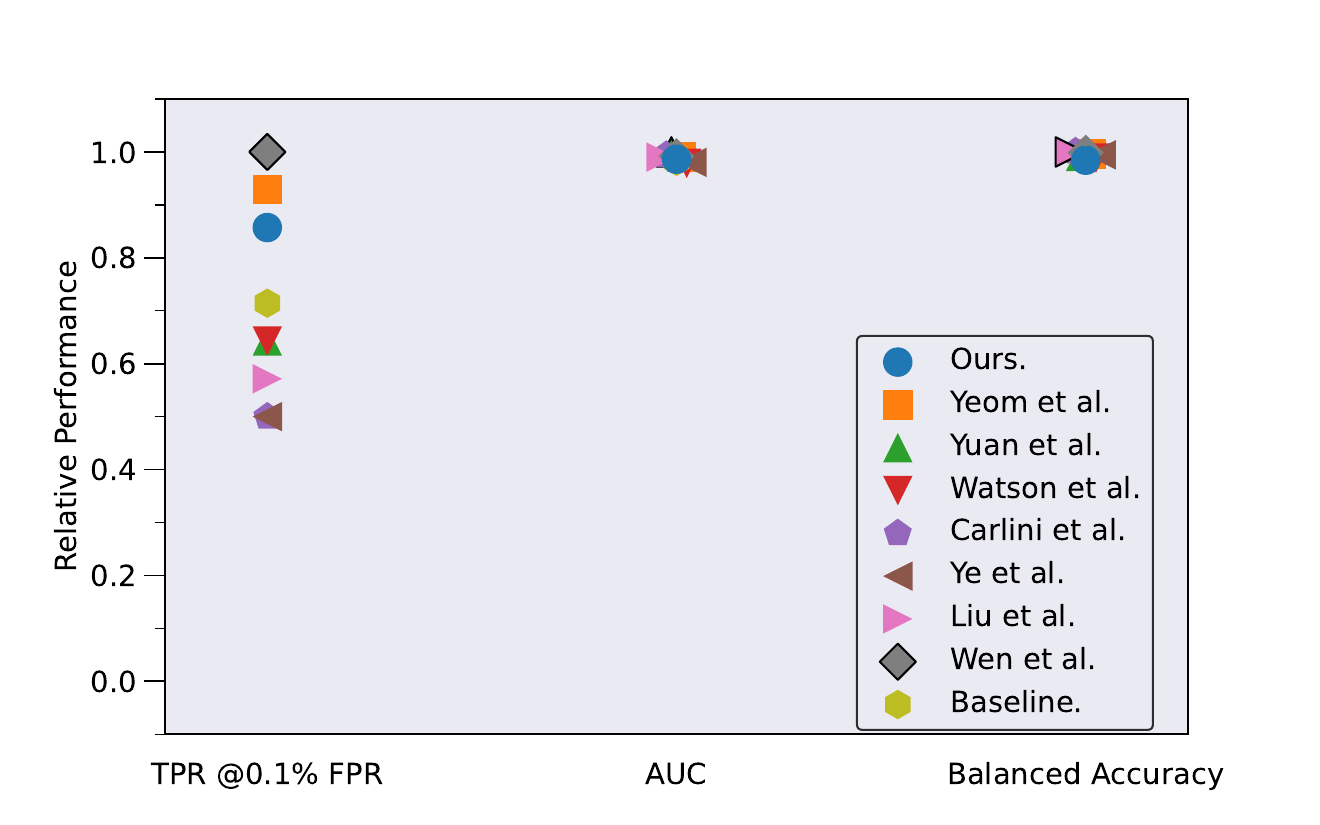} }
    \caption{Attack performance of prior works and our attack against a DenseNet121 model trained on CIFAR-10 using DP-SGD. The baseline employs random guessing to identify members.}
\label{Figure 16}
\end{figure*}

\clearpage
\section{Additional Experimental Results on Cb and Cola Datasets}
\label{Additional Experimental Results on cola and cb}
The notably higher TPR results of the attack on BERT-large compared to BERT-base in Table \ref{table:attack on cb} require further investigation, and we suspect it may be related to the small size of cb (more susceptible to randomness), the special distribution properties of cb, or the more powerful memorization capability of BERT-large.

\begin{table*}[!h]
\centering
\setlength{\tabcolsep}{2.5pt}
\caption{The attack results of BERT-base and BERT-large models trained on cb. Please note that here we consider FPRs down to 0.0 because the cb dataset is very small, making it unable to set FPRs at 0.1\%.}
\scalebox{1.0}
{
\begin{tabular}{l|cccccc}
\toprule
 & \multicolumn{2}{c}{TPR @ 0.0\% FPR}&  \multicolumn{2}{c}{AUC}& \multicolumn{2}{c}{Balanced Accuracy}\\
\cmidrule(l{5pt}r{5pt}){2-3}\cmidrule(l{5pt}r{5pt}){4-5}\cmidrule(l{5pt}r{0pt}){6-7}
Attack Method& BERT-base& BERT-large& BERT-base& BERT-large& BERT-base& BERT-large\\
\midrule
Duan et al.~\cite{duan2023privacy}&0.0\%&5.9\%&0.711&0.628&64.7\%&\textbf{62.7}\%\\
Waston et al.~\cite{watson2021importance}&0.0\%&5.9\%&0.782&0.639&60.8\%&61.8\%\\
\midrule
Ours&\textbf{3.9\%}&\textbf{7.8\%}&\textbf{0.803}&\textbf{0.666}&\textbf{65.7\%}&\textbf{62.7\%}\\
\bottomrule
\end{tabular}
}
\label{table:attack on cb}
\end{table*}

\begin{table*}[!h]
\centering
\setlength{\tabcolsep}{2.5pt}
\caption{The attack results of BERT-base and BERT-large models trained on cola.}
\scalebox{1.0}
{
\begin{tabular}{l|cccccc}
\toprule
 & \multicolumn{2}{c}{TPR @ 0.1\% FPR}&  \multicolumn{2}{c}{AUC}& \multicolumn{2}{c}{Balanced Accuracy}\\
\cmidrule(l{5pt}r{5pt}){2-3}\cmidrule(l{5pt}r{5pt}){4-5}\cmidrule(l{5pt}r{0pt}){6-7}
Attack Method& BERT-base& BERT-large& BERT-base& BERT-large& BERT-base& BERT-large\\
\midrule
Duan et al.~\cite{duan2023privacy}&0.1\%&0.0\%&0.634&0.592&62.3\%&\textbf{62.2}\%\\
Waston et al.~\cite{watson2021importance}&0.3\%&\textbf{0.1}\%&0.656&0.677&59.1\%&61.7\%\\
\midrule
Ours&\textbf{0.5\%}&\textbf{0.1\%}&\textbf{0.701}&\textbf{0.687}&\textbf{62.5\%}&62.1\%\\
\bottomrule
\end{tabular}
}
\label{table:attack on cola}
\end{table*}

\end{document}